\documentclass[12pt]{article}
\usepackage{lmodern}
\usepackage[english]{babel}
\usepackage{pdflscape, pslatex} 
\usepackage{epstopdf}
\usepackage{float}
\usepackage{bm}
\usepackage[authoryear]{natbib}
\usepackage{times}
\usepackage{adjustbox} 
\usepackage{rotating}
\usepackage{setspace}
\usepackage{listings} 
\usepackage{array}
\usepackage{multirow, multicol} 
\usepackage{tabulary,ctable,tabularx}
\usepackage{booktabs}
\usepackage{graphicx}
\usepackage[toc,page]{appendix}
\usepackage{enumerate}
\usepackage{times}
\usepackage{subfig}
\usepackage{amsmath,amsfonts,amssymb,amsthm,titling}

\usepackage[a4paper,width=160mm,top=30mm,bottom=30mm]{geometry}

\usepackage{authblk}

\date{}

\newcommand{\blind}{0}

\begin{document}
	\def\spacingset#1{\renewcommand{\baselinestretch}%
		{#1}\small\normalsize} \spacingset{1}
\if0\blind
{
	\title{\bf A Simulation Comparison of Estimators of Conditional Extreme Value Index under Right Random Censoring}
	\author{Richard Minkah\thanks{
			The authors gratefully acknowledge the University of Ghana-Carnegie Cooperation Next Generation of African Academics for providing financial Assistance.}\hspace{.2cm}\\
		Department of Statistics and Actuarial Science, University of Ghana, Ghana\\
		and \\
		Tertius de Wet  \\
		Department of Statistics and Actuarial Science, Stellenbosch University, South Africa\\
		and\\
		Ezekiel N. N. Nortey\\
		Department of Statistics and Actuarial Science, University of Ghana, Ghana
	}
	\maketitle
} \fi

\bigskip	
\begin{abstract}
	In extreme value analysis, the extreme value index plays a vital role as it determines the tail heaviness of the underlying distribution and is the primary parameter required for the estimation of other extreme events. In this paper, we review the estimation of the extreme value index when observations are subject to right random censoring and the presence of covariate information. In addition, we propose some estimators of the extreme value index, including a maximum likelihood estimator from a perturbed Pareto distribution. The existing estimators and the proposed ones are compared through a simulation study under identical conditions. The results show that the performance of the estimators depend on the percentage of censoring, the underlying distribution, the size of extreme value index and the number of top order statistics. Overall, we found the proposed estimator from the perturbed Pareto distribution to be robust to censoring, size of the extreme value index and the number of top order statistics. 	
\end{abstract}
\noindent%
\vfill
\section{Introduction}

The study of extreme events has received much attention in many fields of application due to the nature of their impact. For instance, extreme earth quakes cause many deaths and destruction to properties; large price movements in equities result in huge losses, profits or collapse of financial markets; large insurance claims lead to solvency problems.

Unlike traditional statistical methods that focus on the central part of distributions, statistics of extremes focuses on the tail of the underlying distribution. Interest is then on parameters associated with the tail of the underlying distribution, such as high quantiles and exceedance probabilities. For inference on such parameters, distributional results are needed on the extreme observations. The first such result in extreme value theory was obtained by \cite{Fisher1928} and further developed by e.g. \citet{Gnedenko1943} and \citet{deHaan1970}. These asymptotic distributions form the basis for carrying out inference in extreme value analysis. 

Let $Y_1,\ldots,Y_n$ be an independent and identically distributed sample on some random variable $Y$ and let  $Y_{1,n}\le\ldots\le Y_{n,n}$ be the corresponding order statistics. The mentioned results state that, if there exists normalising constants $a_n>0$ and $b_n\in\mathbb{R},$ and some nondegenerate function $\Psi,$ such that

\begin{equation}\label{MaxLim}
\frac{Y_{n,n}-b_n}{a_n}\overset{d}{\longrightarrow}\Psi,
\end{equation}
then the constants can be redefined such that for $\gamma\in\mathbb{R},$

\begin{equation} \label{GEV}
\Psi(y)\equiv\Psi_\gamma(y)=\left\{\begin{array}{ll}
\exp\left(-\left(1+\gamma y\right)^{-1/\gamma}\right), & 1+\gamma y>0,~ \gamma\ne 0,\\
\exp\left(-\exp\left(y\right)\right), & y\in\mathbb{R}, ~\gamma=0.\end{array} \right.
\end{equation}
 Here, (\ref{GEV}) is the so-called Generalised 
 Extreme Value distribution and $\gamma$ is the extreme value index (or tail index). The parameter $\gamma$ is the primary parameter needed in extreme value analysis and determines the tail heaviness of the underlying distribution. If $\gamma>0, \Psi_\gamma$ belongs to the Pareto domain (heavy tailed); if $\gamma<0, \Psi_\gamma$ belongs to the Weibull domain (short-tailed); and if $\gamma=0, \Psi_\gamma$ belongs to the Gumbel domain (light-tailed). An underlying distribution function, $F,$ for which (\ref{MaxLim}) and (\ref{GEV}) hold, is said to be in the domain of attraction of $\Psi_\gamma,$ denoted $F\in D\left(\Psi_\gamma\right).$
 
 The estimation of $\gamma$ has been addressed in many papers, including \citet{Hill1975,deHaan1998,Panaretos2003,Beirlant2004,Gomes2008,Gomes2014}. When covariate information is available, the focus is to include it in the estimation by modelling the parameters of the extreme value distribution as a function of the covariate(s). For example, \citet{Davison1990} fitted a Generalised Pareto (GP) distribution with parameters taken as an exponential function of the covariates; \citet{Gardes2008} used moving-window methodology; \citet{Beirlant2003} and \citet{Wang2009} used a conditional exponential regression model; and \citet{Beirlant2004b} employed repeated fitting of local polynomial maximum likelihood estimation. 
 
 In the case of censoring, \citet{Beirlant2007} and \citet{Einmahl2008} proposed an inverse probability-of-censoring weighted method to adapt classical extreme value index estimators to censoring. Similarly, \citet{Gomes2011} and \citet{Brahimi2013} used this idea to adapt various estimators to censoring. In addition, \citet{Beirlant2010} addressed the issue of censoring, obtaining maximum likelihood estimators by adapting the likelihood function of the generalised Pareto distribution to censoring. Also, \citet{Worms2014} considered estimators based on Kaplan-Meier integration and censored regression. Furthermore, \citet{Ameraoui2016} estimated the extreme value index from a Bayesian perspectives and \citet{Beirlant2017} proposed a reduced-bias estimator based on an extended Pareto distribution. 
 
In the case of the presence of both covariate information and censoring, \citet{Ndao2014} proposed three estimators for the estimation of the conditional extreme value index and extreme quantiles for heavy-tailed distributions. In particular, the Hill, generalised Hill and moment type estimators were proposed using the moving window method \citep{Gardes2008} and adapting the estimators to censoring by utilising the inverse probability-of-censoring weighted method \citep{Beirlant2007,Einmahl2008}. Unlike \citet{Ndao2014}, \citet{Stupfler2016} proposed a moment estimator valid for all domains of attraction. In addition, \citet{Ndao2016} addressed the  estimation of the extreme value index under censoring and the presence of random covariates. 
 
Although quite a number of papers have compared the available estimators of the extreme value index, no paper has compared the estimators of the conditional extreme value index when observations are subject to random censoring. The aim of this paper is two-fold. Firstly, to adapt some classical estimators to the current context, including a reduced-bias maximum likelihood estimator based on a perturbed Pareto distribution in \citet{Beirlant2004}. Secondly, we review the available estimators that were proposed in the literature for estimating conditional extreme value index under right random censoring and compare them together with the proposed ones in a simulation study. 

The remainder of the paper is organised as follows. In Section 2, we set out the framework for the estimation of the parameter of interest i.e. extreme value index. In addition, the existing estimators are reviewed and we present the proposed estimators. In section 3, we conduct a simulation study to assess the performance of these estimators. 
 Lastly, general conclusions from the simulation results are presented in section 4.
\section{Estimation of the Extreme Value Index}

Consider $Y_1, \ldots, Y_n$ as independent copies of a positive random variable, $Y,$ and let $\bm{x}_1, \ldots, \bm{x}_n$ be the values of an associated $d$-dimensional covariate vector, $\bm{x}\in \Omega,$ where $\Omega\subset \mathbb{R}^d.$  Also, in order to incorporate the presence of censoring, let $C_1, \ldots, C_n$ be independent copies of another positive random variable $C,$ also associated with the covariate vector $\bm{x}.$ We assume that for all $\bm{x}\in\Omega,$ the random variables, $Y$ and $C,$ are independent. Furthermore, for every $\bm{x}\in\Omega,$ we assume that the random variables $Y$ and $C$ have respective conditional distribution functions, $F(.;\bm{x})\in D(\Psi_{\gamma_1(\bm{x})})$ and $G(.;\bm{x})\in D(\Psi_{\gamma_2(\bm{x})}),$ where $\gamma_1(\bm{x})$ and $\gamma_2(\bm{x})$ are real functions. We consider the case where $F(.;\bm{x})$ and $G(.;\bm{x})$ are in the Pareto domain of attraction i.e. $\gamma_1(\bm{x})>0$ and $\gamma_2(\bm{x})>0.$ For this domain of attraction, the distribution functions can be represented as
\begin{equation}\label{FtDomain 1-F}
1-F(y;\bm{x})=y^{-\frac{1}{\gamma_1(\bm{x})}}\ell_F(y;\bm{x}) ~\mbox{and}~1-G(y;\bm{x})=y^{-\frac{1}{\gamma_2(\bm{x})}}\ell_G(y;\bm{x}).
\end{equation}
or equivalently in terms of the tail quantile function, 
\begin{equation}
U_F(y;\bm{x})=y^{\gamma_1(\bm{x})}\ell_{U_F}(y;\bm{x}) ~~\mbox{and}~~U_G(y;\bm{x})=y^{\gamma_2(\bm{x})}\ell_{U_G}(y;\bm{x}).
\end{equation}
Here, $\ell_F(y;\bm{x})~~\left(\mbox{and}~~\ell_{U_F}(y;\bm{x})\right)$ and $\ell_G(y;\bm{x})~\left(\mbox{and}~ \ell_{U_G}(y;\bm{x})\right)$ are slowly varying functions associated with $F$ and $G$ respectively and defined as
\begin{equation}\label{s.v}
\lim\limits_{y\to\infty}\frac{\ell_j(by;\bm{x})}{\ell_j(y;\bm{x})}=1,~~b>0; j\in\{F, G,U_F, U_G\}.
\end{equation}

In this context, we observe the triplets $\{Z_i, \delta_i, \bm{x}_i\},~i=1,\ldots,n$ where $Z_i=\min\{Y_i,~C_i\} $ and  $\delta_i=\mathbb{I}{\{Y_i\le C_i\}}.$ By the independent assumption of $Y$ and $C,$ the conditional distribution function $H(.;\bm{x})$ of the random variable $Z_i$ is related to $F(.;\bm{x})$ and $G(.;\bm{x}),$ as 

\begin{equation}
1-H(.;\bm{x})= \left(1-F(.;\bm{x})\right)\left(1-G(.;\bm{x})\right).
\end{equation}
Therefore, $H(.;\bm{x})$ is also in the Pareto domain with conditional extreme value index given in \citet{Einmahl2008} as
\begin{equation}
\gamma(\bm{x})=\frac{\gamma_1(\bm{x})\gamma_2(\bm{x})}{\gamma_1(\bm{x})+\gamma_2(\bm{x})}.
\end{equation}

Before presenting the estimators, we define a ball $B(\bm{x},r)$ in $\Omega$ where  $\bm{x}$ and $r~(r>0)$ are the center and radius respectively. Thus, 

\begin{equation}\label{ball_rad}
B(\bm{x},r)=\{\mu\in \mathbb{R}^d:~ d(\bm{x},\mu)\le r\}.
\end{equation}
In addition, let $h_{n,\bm{x}}$ be a positive integer that approaches $0$ as $n\to \infty$. The estimators of the EVI are based on observations of $Z_i$ for which the corresponding values of $\bm{x}_i$ fall within the ball $B(\bm{x},h_{n,\bm{x}}).$ The proportion of the design points falling within the ball is defined as

\begin{equation} \label{bal}
\phi(h_{n,\bm{x}})=\frac{1}{n}\sum_{i=1}^{n}\mathbb{I}{\{\bm{x}_i\in B(\bm{x},h_{n,\bm{x}})\}},
\end{equation}
where $\mathbb{I}$ is the indicator variable. Relation (\ref{bal}) plays an important role in this procedure as it describes how the points are concentrated around the neighbourhood of $\bm{x}_i$ when $h_{n,\bm{x}}$ approaches $0$ \citep{Gardes2008}. The number of nonrandom observations in $(0,\infty)\times B(\bm{x},h_{n,\bm{x}})$ is given by $m_{n,\bm{x}}=n\phi(h_{n,\bm{x}})$.

Let $\left(W_1(\bm{x}),\delta_{(1)}\right), \ldots, (W_{m_{n,\bm{x}}}(\bm{x}),\delta_{(m_{n,\bm{x}})})$ denote the pair, $\left(Z_i,\delta_i\right), i=1,2,\ldots,n,$ that have their corresponding $\bm{x}_i$-values falling within the ball as defined in (\ref{ball_rad}). Also, let  $W_{1,m_{n,\bm{x}}}(\bm{x}) \le\ldots\le W_{m_{n,\bm{x}},m_{n,\bm{x}}}(\bm{x})$ be the corresponding order statistics of $W$'s and $\delta_{(i)}^{(W)},~i=1,\ldots,m_{n,\bm{x}}$ be the values of $\delta$'s associated with $W_{i,m_{n,\bm{x}}}(\bm{x}), i=1,\ldots,m_{n,\bm{x}}.$  The values of $\delta_{(i)}^{(W)},~i=1,\ldots,m_{n,\bm{x}}$ form the basis for adapting the classical estimators of the conditional extreme value index presented below to censoring. 

In what follows, given a sample $\{Z_1, \delta_1, \bm{x}_1\},\ldots,\{Z_n, \delta_n, \bm{x}_n\},$ we consider the estimation of $\gamma_1(\bm{x}).$ To do this, we rely on the observations $\left(W_1(\bm{x}),\delta_{(1)}\right), \ldots, (W_{m_{n,\bm{x}}}(\bm{x}),\delta_{(m_{n,\bm{x}})})$ resulting from the moving window approach of \citet{Gardes2008} described after (\ref{bal}).

\subsection{The Existing Estimators}

The existing estimators result from the application of the moving window technique \citep{Gardes2008} and the inverse probability-of-censoring weighted method \citep{Beirlant2007,Einmahl2008} to adapt classical estimators to censoring. \citet{Ndao2014} used this approach to adapt the Hill, generalised Hill and moment estimators to censoring. In this section, we review these estimators and follow a similar approach to propose other estimators of the extreme value index in the next section.

The estimators introduced by \citet{Ndao2014} are presented are the following:

\begin{itemize}
	\item \textbf{The Hill-type estimator}: The Hill estimator \citep{Hill1975} is arguably the most common estimator of $\gamma$ in the Pareto case i.e. $\gamma>0.$ To take into account the available covariate information, the Hill estimator is defined for the $(k_{n,\bm{x}}+1)$-largest order statistics as
	
	\begin{equation} \label{Hill_6}
	\hat{\gamma}^{(c,Hill)}(W,k_{n,\bm{x}},m_{n,\bm{x}})=\frac{1}{k_{n,\bm{x}}}\sum_{i=1}^{k_{n,\bm{x}}}i\left(\log{W_{m_{n,\bm{x}}-i+1,m_{n,\bm{x}}}(\bm{x})-\log{W_{m_{n,\bm{x}}-i,m_{n,\bm{x}}}(\bm{x})}}\right).
	\end{equation}

	\item \textbf{The Moment-type estimator}:  \citet{Dekkers1989} introduced  the moment estimator as an adaptation of the Hill estimator valid for all domains of attraction. It is defined to take into account the covariate information $\bm{x}$ as,
	\begin{equation}\label{MOM_6}
	\hat{\gamma}^{(c,MOM)}(W,k_{n,\bm{x}},m_{n,\bm{x}})=M^{(1)}_n(W,k_{n,\bm{x}},m_{n,\bm{x}})+1-\frac{1}{2}\left(1-\frac{\left(M^{(1)}_n(W,k_{n,\bm{x}},m_{n,\bm{x}})\right)^2}{M^{(2)}_n(W,k_{n,\bm{x}},m_{n,\bm{x}})} \right)^{-1}
	\end{equation}
	where
	
	\begin{equation}\label{Mj_6}
	M^{(j)}_n(W,k_{n,\bm{x}},m_{n,\bm{x}})=\frac{1}{k_{n,\bm{x}}}\sum_{i=1}^{k_{n,\bm{x}}}
	\left[\log{(W_{m_{n,\bm{x}}-i+1,m_{n,\bm{x}}}(\bm{x}))}-\log{(W_{m_{n,\bm{x}}-k_{n,\bm{x}},m_{n,\bm{x}}}(\bm{x}))}\right]^j.
	\end{equation}

	\item \textbf{The generalised Hill Estimator}:  \citet{Beirlant1996} proposed the generalised Hill (GH) estimator as an attempt to extend the Hill estimator to the case where $\gamma \in \mathbb{R}.$ The GH estimator is obtained as the slope of the ultimately linear part of the generalised Pareto quantile plot of the observations within the defined window as, 
	
	\begin{equation}\label{UH_6}
	\hat{\gamma}^{(c,UH)}(W,k_{n,\bm{x}},m_{n,\bm{x}})= \frac{1}{k_{n,\bm{x}}}\sum_{j=1}^{k_{n,\bm{x}}}\log{UH}_{j,m_{n,\bm{x}}}-\log UH_{k_{n,\bm{x}}+1,m_{n,\bm{x}}}, 
	\end{equation}
	where \[UH_{j,m_{n,\bm{x}}}=W_{m_{n,\bm{x}}-j,m_{n,\bm{x}}}\left( \frac{1}{j}\sum_{i=1}^{j}\log{W_{m_{n,\bm{x}}-i+1,m_{n,\bm{x}}}}-\log{W_{m_{n,\bm{x}}-j,m_{n,\bm{x}}}}\right).\]
	
\end{itemize}

The estimators (\ref{Hill_6}), (\ref{MOM_6}) and (\ref{UH_6}) were adapted to censoring by dividing each estimator by the proportion of noncensored observations in the $k_{n,\bm{x}}$ largest order statistics of $W$'s. Thus, an adapted estimator of the conditional extreme value index is given by

\begin{equation}\label{adapt}
\hat{\gamma}^{(c,.)}({W,k_{n,\bm{x}},m_{n,\bm{x}}})=\frac{\hat{\gamma}^{(.)}_{W,k_{n,\bm{x}},m_{n,\bm{x}}}}{\hat{\wp}(\bm{x})},
\end{equation}

where $\hat{\wp}(\bm{x})=k_{n,\bm{x}}^{-1}\sum_{i=1}^{k_{n,\bm{x}}}\delta_{m_{n,\bm{x}}-i+1,m_{n,\bm{x}}}^{(w)}.$

\subsection{The Proposed Estimators}
In this section, we propose some estimators for the estimation of conditional extreme value index when observations are subject to right random censoring. We follow closely the methodology in \citet{Ndao2014}. 

Firstly, the estimators are presented to take into account the covariate and are subsequently adapted to censoring. 

\begin{enumerate}
	\item[(i)]  \textbf{The Zipf estimator}:  \cite{Kratz1996} derived the Zipf estimator as a smoother version of the Hill estimator through unconstrained least squares fit to the $k$ largest observations on the generalised Pareto quantile plot method of \cite{Beirlant1996}. The estimator is valid for $\gamma>0$ and in the case of a covariate, is given by, 	
	\begin{equation} \label{zipf_6}
	\hat{\gamma}^{(c,Zipf)}(W,k_{n,\bm{x}},m_{n,\bm{x}})=\sum_{i=1}^{k_{n,\bm{x}}}i\log{\left(\frac{W_{m_{n,\bm{x}}-i+1,m_{n,\bm{x}}}(\bm{x})}{W_{m_{n,\bm{x}}-i,m_{n,\bm{x}}}(\bm{x})}\right)}\log(k_{n,\bm{x}}/i)\Bigg/\sum_{i=1}^{k_{n,\bm{x}}}\log{(k_{n,\bm{x}}/i)},
	\end{equation}

	\item[(ii)] \textbf{The Moment Ratio}: The Moment Ratio estimator was introduced by \citet{Danielsson1996} as a moment based estimator to  reduce bias in the Hill estimator. The moment ratio estimator is  valid for the Pareto domain of attraction only. In the case of a covariate, it is given by
		
	\begin{equation}\label{MoMR_6}
	\hat{\gamma}^{(MomR)}({W,k_{n,\bm{x}},m_{n,\bm{x}}})=\frac{1}{2}\frac{M_{W,k_{n,\bm{x}},m_{n,\bm{x}}}^{(2)}}{M_{W,k_{n,\bm{x}},m_{n,\bm{x}}}^{(1)}},
	\end{equation}
	where $M_{W,k_{n,\bm{x}},m_{n,\bm{x}}}^{(j)}, ~j=1,2$ is obtained from (\ref{Mj_6}).

	\item[(iii)] \textbf{The Peng Moment Estimator}: \citet{Deheuvels1997} reports on a variant of the moment estimator for the no covariate case suggested by Liam Peng. This estimator is designed to reduce bias in the moment estimator and it is adapted to the covariate case as 
	
	\begin{equation}\label{PMoM_6}
	\hat{\gamma}^{(PMom)}({W,k_{n,\bm{x}},m_{n,\bm{x}}})= \frac{1}{2}\frac{M_{W,k_{n,\bm{x}},m_{n,\bm{x}}}^{(2)}}{M_{W,k_{n,\bm{x}},m_{n,\bm{x}}}^{(1)}}+1-\frac{1}{2}\left(1-\frac{(M_{W,k_{n,\bm{x}},m_{n,\bm{x}}}^{(1)})^2}{M_{W,k_{n,\bm{x}},m_{n,\bm{x}}}^{(2)}}\right)^{-1},
	\end{equation}
	where $M_{W,k_{n,\bm{x}},m_{n,\bm{x}}}^{(j)}, ~j=1,2$ is as before obtained from (\ref{Mj_6}). This estimator is valid for all domains of attraction.
	
	\item[(iv)]  \textbf{The Perturbed Pareto Estimator}

	\cite{Beirlant2004} derived the perturbed Pareto estimator as a reduced-biased Hill type estimator by making use of the second-order properties on the underlying distribution function, $F.$ Here, we consider adapting this estimator to censoring and the presence of covariate information. Considering (\ref{FtDomain 1-F}) and (\ref{s.v}), we can write 
	
	\begin{equation}\label{DM_1-F(xt)/1-F}
	\lim\limits_{u\to \infty}\frac{1-F(uw;\bm{x})}{1-F(u;\bm{x})}=w^{-1/{\gamma(\bm{x})}},~\mbox{ for any} ~w>1.
	\end{equation}
	Therefore, (\ref{DM_1-F(xt)/1-F}) can be interpreted as
	
	\begin{equation}\label{P(X/u|X>u)}
	1-F_u(w;\bm{x})=P(W/u>z|W>u)\approx w^{-1/\gamma(\bm{x})},
	\end{equation}
	for large $u$ and $w>1.$ Now, consider the relative excesses $V_j=W_i/u~ \mbox{given} ~W_i>u$ for a large threshold $u,$ where $i$ is the index of the $j$th exceedance. Then, it seems natural to consider the strict Pareto distribution as the approximate distribution of the relative excesses, $V_j.$ The maximum likelihood estimator of the parameter results in the Hill estimator in (\ref{Hill_6}). However, if the strict Pareto approximation is poor, the Hill estimator has large bias, and hence, a second-order refinement is needed to address the departure from the strict Pareto distribution \citep[see][]{Beirlant2004}. Assume that the $s.v$ function, $\ell,$ satisfies the second-order assumption:\\  
	
	\textbf{Assumption I: }	There exists a real constant $\rho<0$ and a rate function $b$ satisfying $b(w)\to 0$ as $w\to \infty,$ such that for all $\lambda\geq 1$,

	\begin{equation}\label{2nd s.v}
	\lim\limits_{w\to \infty}\frac{\log\ell(\lambda w)-\log\ell(w)}{b(w)}=\kappa_\rho(\lambda)
	\end{equation}
	where $\kappa_\rho(\lambda)=\int_{1}^{\lambda}u^{\rho-1}du$ \citep[page  602]{Beirlant1999a}.\\
	
Then, from (\ref{2nd s.v}), we can write  (\ref{DM_1-F(xt)/1-F}) as
	
	\begin{equation}\label{1-F(U)PPD}
	\lim\limits_{u\to \infty}\frac{1-F(uw;\bm{x})}{1-F(u;\bm{x})}=w^{-1/\gamma(\bm{x})}\left(1-\frac{b(u)}{\tau(\bm{x})}\left(w^{-\tau(\bm{x})}-1\right)+o\left(b\left(u\right)\right)\right),~ \tau(\bm{x})>0,
	\end{equation}
	where $b$ is regularly varying with index -$\tau(x).$ Ignoring the error term, (\ref{1-F(U)PPD}) becomes a mixture of two Pareto distributions. The survival function for such a distribution is given by
	\begin{equation}\label{PPD_DF}
	1-G(w;\bm{x})=\left(1-c(\bm{x})\right)w^{-1/\gamma(\bm{x})}+c(\bm{x})w^{-\left(1/\gamma(\bm{x})+\tau(\bm{x})\right)}
	\end{equation}
	where $c(\bm{x})\in (-1/\tau(\bm{x}), 1), ~ \tau(\bm{x})>0$ and $w>1.$ In practice, the perturbed Pareto distribution is fitted to the relative excesses, $V_j, ~ j=1, ..., k_{n,\bm{x}},$ and the parameters of the distribution can be estimated through the maximum likelihood method.  The resulting estimator is denoted by $\hat{\gamma}^{(PPD)}\left({V,k_{n,\bm{x}},m_{n,\bm{x}}}\right).$ 
Similar to \citet{Ndao2014}, the estimators of the the extreme value index from (i) through to (iv) are adapted to censoring using (\ref{adapt}).

Furthermore, we extend the two estimators of the extreme value index introduced in \citet{Worms2014} when observations are subject to random censoring to the case where covariate information is available. 

\item[(v)] The first estimator is given by

\small
\begin{equation}\label{WW.KL_6}
\hat{\gamma}^{(c, WW.KM)}({W,k_{n,\bm{x}},m_{n,\bm{x}}}) := \frac{1}{n\left(1-\hat{F}\left(W_{m_{n,\bm{x}}-k_{n,\bm{x}},m_{n,\bm{x}}}\right)\right)}\sum_{j=1}^{k_{n,\bm{x}}}\frac{\delta_{m_{n,\bm{x}}-j+1,m_{n,\bm{x}}}}{1-\hat{G}(W_{m_{n,\bm{x}}-j+1,m_{n,\bm{x}}}^-)}\log{\left(\frac{W_{m_{n,\bm{x}}-j+1,m_{n,\bm{x}}}}{W_{m_{n,\bm{x}}-k_{n,\bm{x}},m_{n,\bm{x}}}}\right)},
\end{equation}
\normalfont
where $\hat{F}$ and $\hat{G}$ are respectively the Kaplan-Meier estimators for $F$ and $G$ given by 

\begin{equation}
1-\hat{F}(b)= \Pi_{W_{j,m_{n,\bm{x}}}\leq b} \left( \frac{m_{n,\bm{x}}-j}{m_{n,\bm{x}}-j+1}\right)^{\delta_{j,m_{n,\bm{x}}}}
\end{equation}
and
\begin{equation}\label{1-G(Z^-)_6}
1-\hat{G}(b)= \Pi_{W_{j,m_{n,\bm{x}}}\leq b} \left( \frac{m_{n,\bm{x}}-j}{m_{n,\bm{x}}-j+1}\right)^{1-\delta_{j,m_{n,\bm{x}}}},
\end{equation}
for $b<W_{m_{n,\bm{x}},m_{n,\bm{x}}}.~$ Here, $\hat{G}\left(W_{m_{n,\bm{x}}-j+1,m_{n,\bm{x}}}^-\right)~$  is defined as a function of the form $g(w^-)=\lim\limits_{\nu\to w}g(\nu).$ 

\item[(vi)] The second alternative estimator is a weighted version of the Hill-type estimator (\ref{WW.KL_6}),
\small
\begin{equation}\label{WW.L_6}
\hat{\gamma}^{(c, WW.KL)}({W,k_{n,\bm{x}},m_{n,\bm{x}}}) :=\\ \frac{1}{n\left(1-\hat{F}\left(W_{m_{n,\bm{x}}-k_{n,\bm{x}},m_{n,\bm{x}}}\right)\right)}\sum_{j=1}^{k_{n,\bm{x}}}\frac{\delta_{m_{n,\bm{x}}-j+1,m_{n,\bm{x}}}}{1-\hat{G}(W_{m_{n,\bm{x}}-j+1,m_{n,\bm{x}}}^-)}j\log{\left(\frac{W_{m_{n,\bm{x}}-j+1,m_{n,\bm{x}}}}{W_{m_{n,\bm{x}}-k_{n,\bm{x}},m_{n,\bm{x}}}}\right)}.
\end{equation}

\end{enumerate}

In the next section, the performance of the existing and proposed estimators will be compared via a simulation study.

\section{Simulation Study}

In this section, we present a simulation study to assess the performance of the estimators discussed in the previous section as the asymptotic distribution of most of the estimators are unknown. 

\subsection{Design}
Several sample sizes were considered in the simulation and for simplicity, we report on the simulation for samples of size $n=2000$ generated from three distributions i.e. Burr, Fr\'{e}chet and Pareto only. For each distribution, our interest is in the estimation of the conditional extreme value index (EVI) function,  $\gamma_1(x)=e^{\left(\beta_0+\beta_1x\right)}.$ Here, $\beta_0$ and $\beta_1$ were chosen as $-0.11$ and $-2.90$ such that the values of $\gamma_1(x)$ for $x\in \mbox{Uniform}(0,1)$ are within (0,1). This range of values of the extreme value index is the most common in extreme value theory literature for simulation studies and practical applications \citep[see for e.g.][]{Gilli2006,Gomes2011,Einmahl2008,Stupfler2016}. In particular, we selected values of $x$ equal to, $0.12,$  $0.37$ and $0.75,$ corresponding respectively to $\gamma_1(x)$ values, 0.63 (large), 0.31 (medium) and 0.10 (small). The choice of parameter functions for each distribution are presented in Table \ref{Dist6_tab}. 

\begin{table}[htp!]
	\centering
	\caption{ Distributions with parameters as a function of $x$}
	\label{Dist6_tab}
	\begin{tabular}{llllll}
		\toprule       &     &      \multicolumn{4}{c}{\textbf{Parameter function}}\\\cmidrule{3-6}
		\textbf{Distribution} & $1-F(y;x)$   &$\tau(x)$       & $\lambda(x)$        &$\alpha(x)$  & $\gamma_1(x)$\\\hline 
		& & & & & \\
		Burr       & $\left(\frac{\eta(x)}{\eta(x)+y^{\tau(x)}}\right)^{\lambda(x)}$    & $2$             & $0.5e^{\left(0.11+2.90x\right)}$ &  NA &  $e^{-\left(0.11+2.90x\right)}$\\
			& & & & & \\ 
		
		Pareto      & $y^{-\alpha(x)}$   &    NA    & $1$       &$e^{\left(0.11+2.90x\right)}$         & $e^{-\left(0.11+2.90x\right)}$ \\
			& & & & & \\ 
		Fr\'{e}chet& $1-\exp\left(-y^{-\alpha(x)}\right)$  & NA    & NA   & $e^{\left(0.11+2.90x\right)}$  &   $e^{-\left(0.11+2.90x\right)}$ \\
		\bottomrule 
	\end{tabular}
	\flushleft
	\textbf{Note:} $\eta(x)$ the scale parameter was taken as 1. Also, the Pareto distribution is a limiting case of the Burr distribution with $\lambda(x)=1.$ 
\end{table}
In addition, the distribution of $C$ is chosen such that  the percentage of censoring in the right tail is 10\%, 35\% and 55\%. The performance measures used for examining the estimators of $\gamma_1(x)$ are Mean Square Error (MSE) and median bias (hereafter referred to as bias). 

The following algorithm was implemented to obtain the performance measures:  

\begin{enumerate}[\hspace{0.5cm}\textbf{A}1.]
	
	
	\item Generate $n  \left(n=2000\right)$ random observations from, $x\sim \mbox{Uniform}(0,1).$  \label{D1}

	\item Generate  $n,$ random samples from the distributions of $Y$ and $C$ with parameters, $\gamma_1(x)$ and $\gamma_2(x),$ respectively. To maintain an approximately equal percentage of censoring in each sample, $\gamma_2$ is chosen as $\gamma_2(x)=\gamma_1(x)\wp(x)/(1-\wp(x)),$ where $\wp(x)$ is the percentage of noncensored observations.\label{D2}
	
	\item Let $Z_i=\min{\{Y_i, C_i\}}$ and $\delta_i=\mathbb{I}{\{Y_i\le C_i\}},~ i=1, \ldots, n$ to obtain the triplets $(Z_i, \delta_i,x_i),~i=1,2,\ldots,n.$ \label{D3}
	
	\item Choose a covariate value of interest, $x^\star\in[0,1],$  window size, $h,$ and obtain the observations $(Z_i, \delta_i),~i=1,2,\ldots,n^\star,$  with its $x_i$ values falling within the window $[x^\star-h,~x^\star+h],$ where $n^\star$ is the number of observations within the window. \label{D4}
	
	\item Compute an estimate of $\gamma_1(x^\star)$ using $\hat{\gamma}_1^{(c,.)}(x^\star),$  at each number of top order statistics $k\in\{5,\ldots,n^\star\}$ for the sample in A4.\label{D5}
	
	
	\item Repeat A\ref{D1}-A\ref{D5} a large number of times, $R~ (R=1000),$ to obtain $\hat{\bm{\gamma}}_1^{(c,.)}(x^\star)=\left(\hat{\gamma}^{(c,.)}_{1,1}(x^\star),\ldots, \hat{\gamma}^{(c,.)}_{1,R}(x^\star)\right)'$ at each $k.$

\item	At each $k$ value, compute the median bias \[\mbox{bias}\left(\hat{\gamma}^{(c,.)}_1(x^\star)\right)=\mbox{median}\left(\hat{\bm{\gamma}}_1^{(c,.)}(x^\star)\right)-\gamma_1(x^\star)\] and  the MSE, 	
		\[MSE\left(\hat{\gamma}_1^{(c,.)}(x^\star)\right)=\frac{1}{R}\sum_{i=1}^{R}\left[\hat{\gamma}_{1,i}^{(c,.)}(x^\star)-\gamma_1(x^\star)\right]^2\]
	
\end{enumerate} 

\subsection{Results and Discussion}

\begin{figure}[htp!]
	\centering
	\subfloat{%
		\includegraphics[height=4.5cm,width=.33\textwidth]{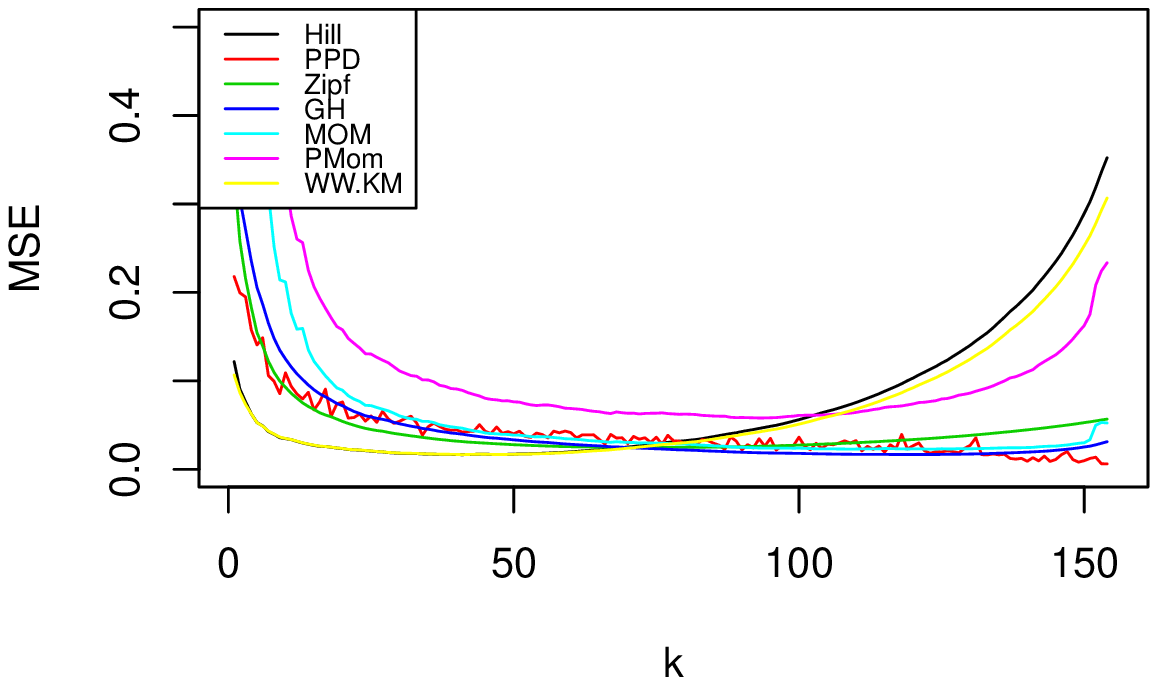}}%
	\subfloat{%
		\includegraphics[height=4.5cm,width=.33\textwidth]{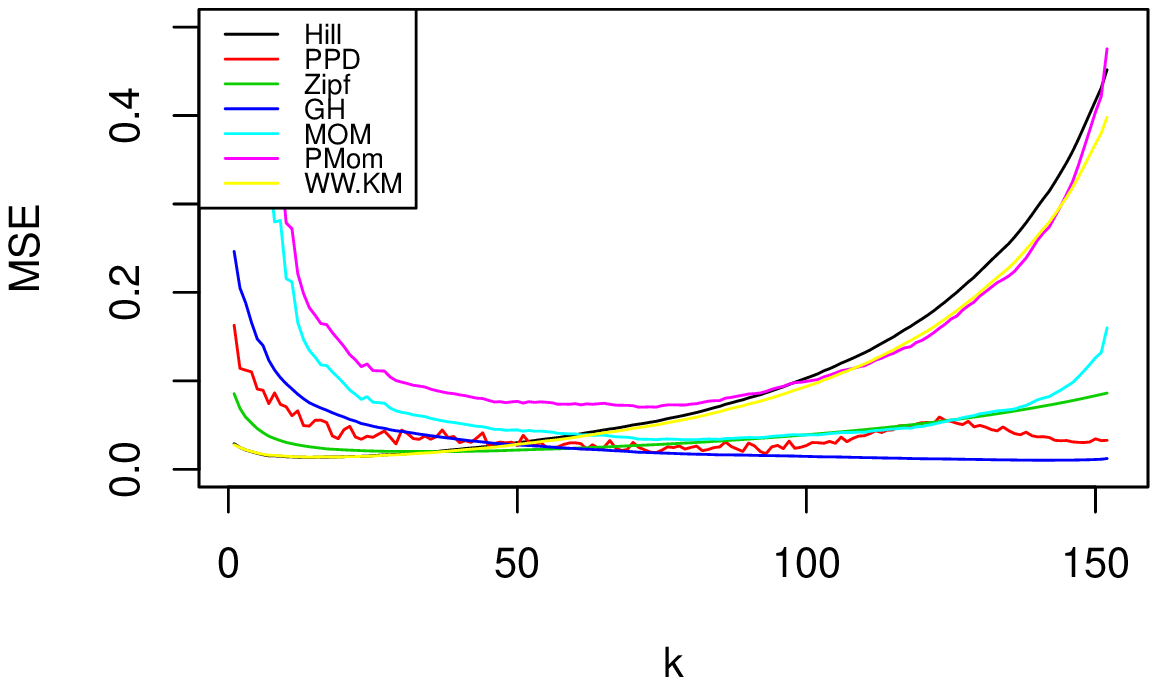}}%
	\subfloat{%
		\includegraphics[height=4.5cm,width=.33\textwidth]{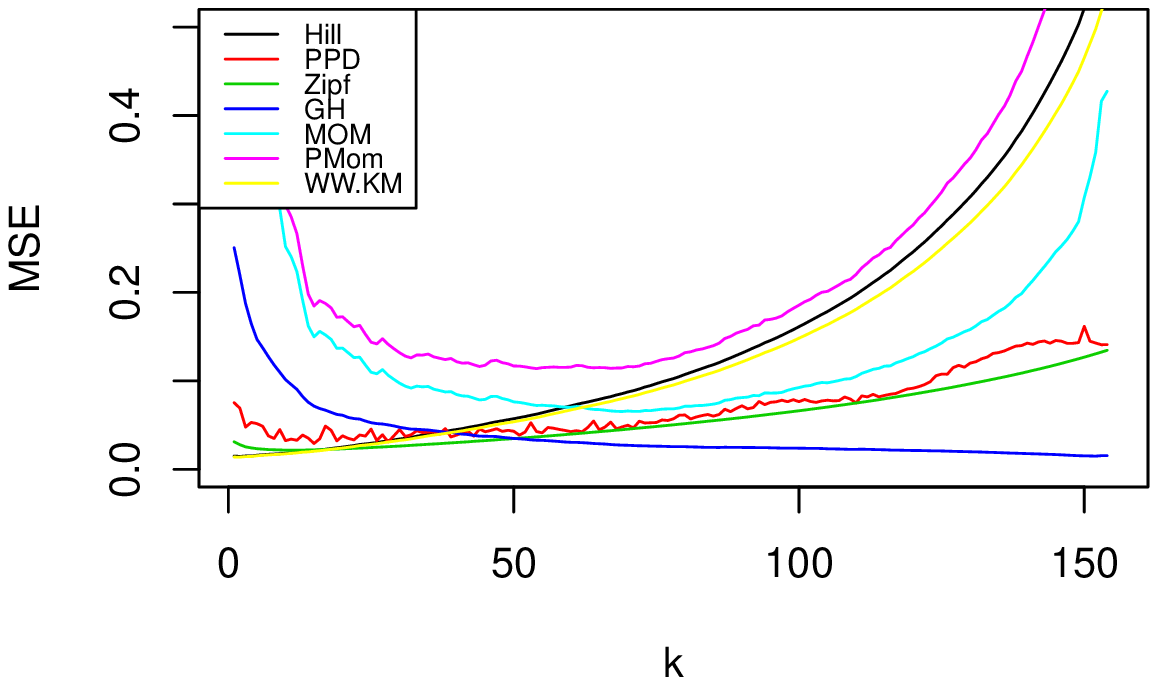}}\\
	\subfloat{%
		\includegraphics[height=4.5cm,width=.33\textwidth]{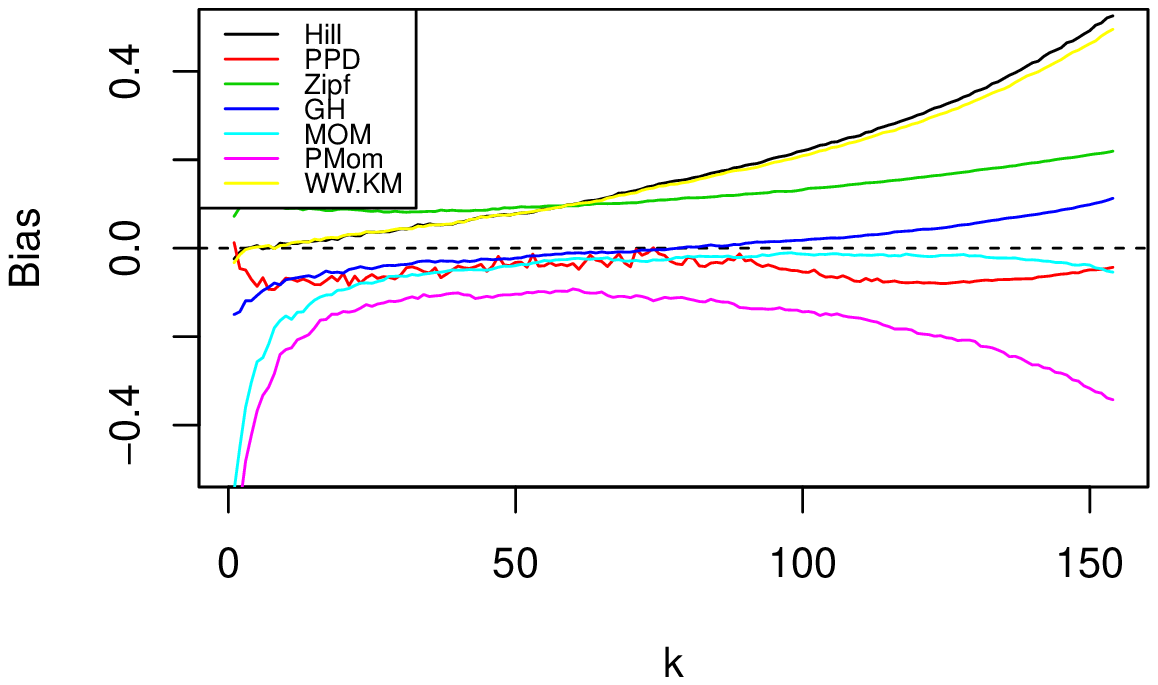}}%
	\subfloat{%
		\includegraphics[height=4.5cm,width=.33\textwidth]{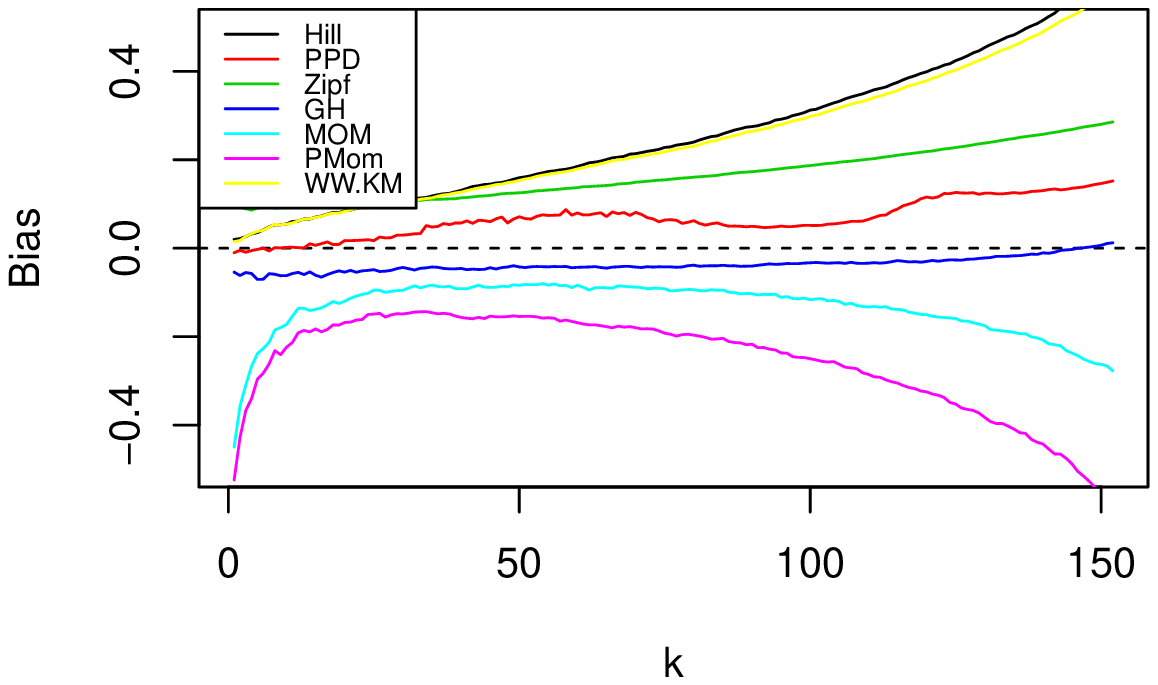}}%
	\subfloat{%
		\includegraphics[height=4.5cm,width=.33\textwidth]{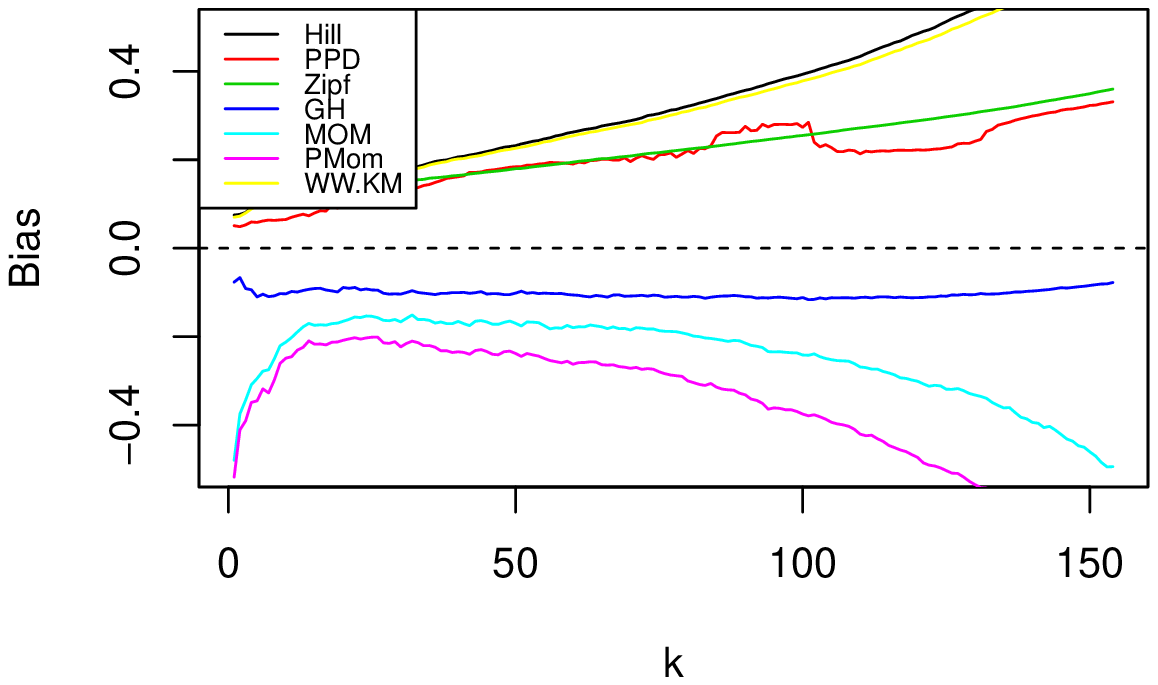}}\\
	
	\caption{ Results for Burr distribution with $\wp=0.1.$  Leftmost column: $\gamma_1(x)=0.63~ (x=0.12);$ Middlemost column : $\gamma_1(x)=0.31~ (x=0.37);$ Rightmost column: $\gamma_1(x)=0.10~ (x=0.75);$}
	\label{B1}
\end{figure}

\begin{figure}[htp!]
	\centering
	\subfloat{%
		\includegraphics[height=4.5cm,width=.33\textwidth]{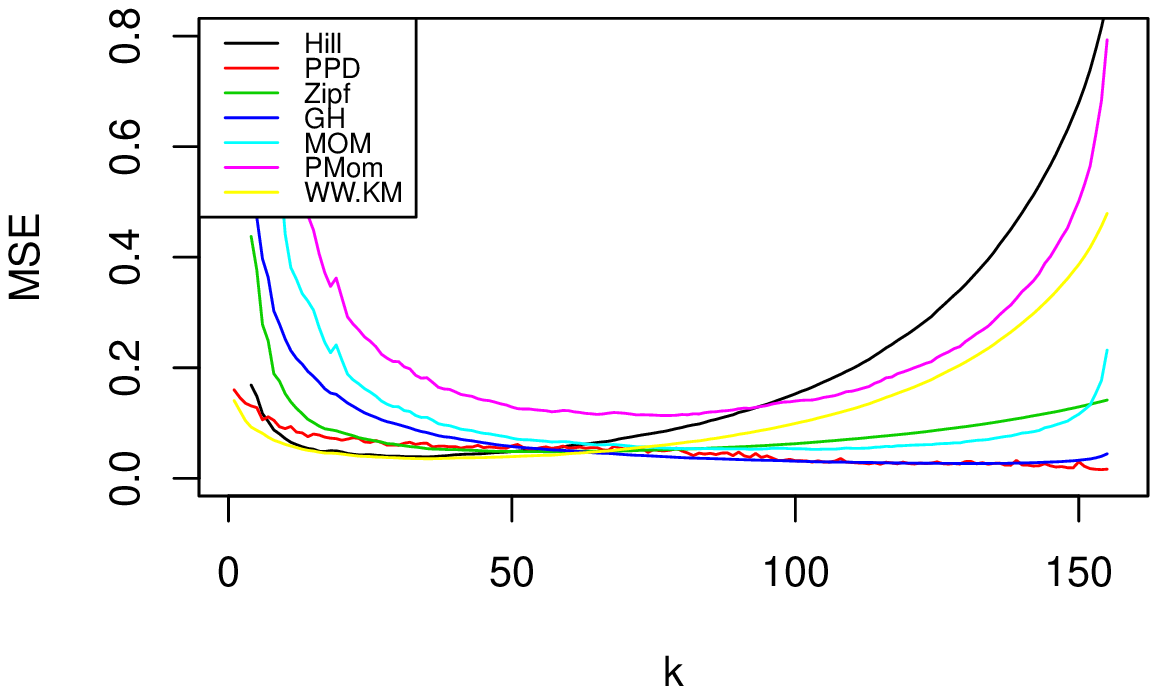}}%
	\subfloat{%
		\includegraphics[height=4.5cm,width=.33\textwidth]{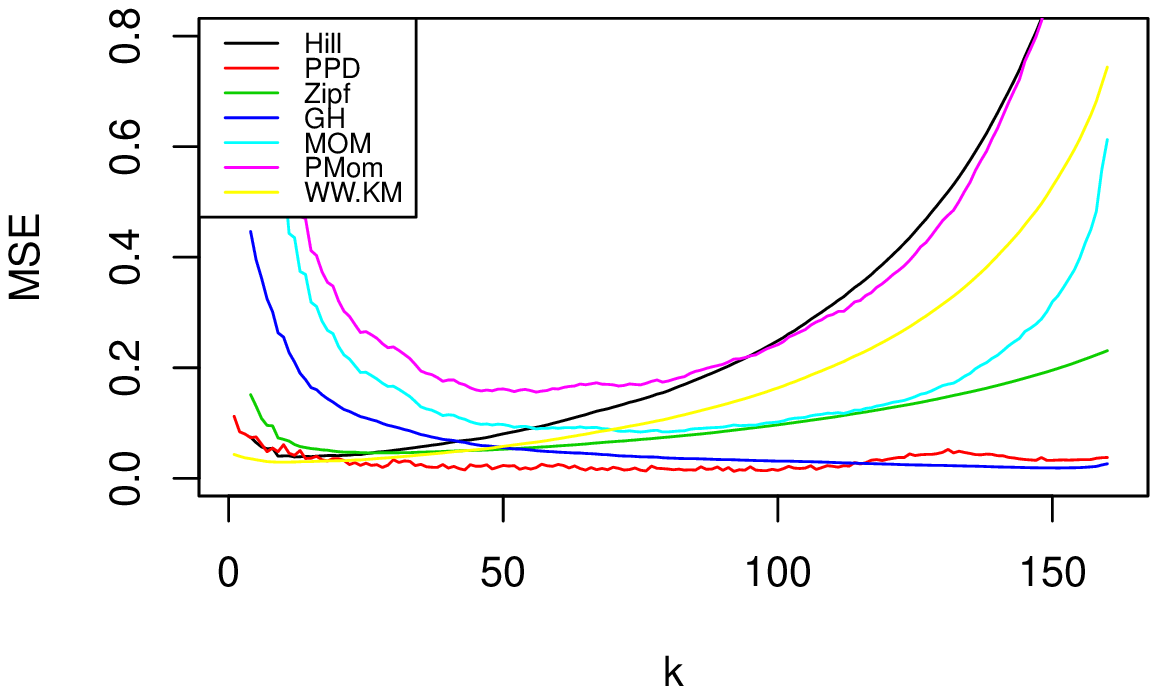}}%
	\subfloat{%
		\includegraphics[height=4.5cm,width=.33\textwidth]{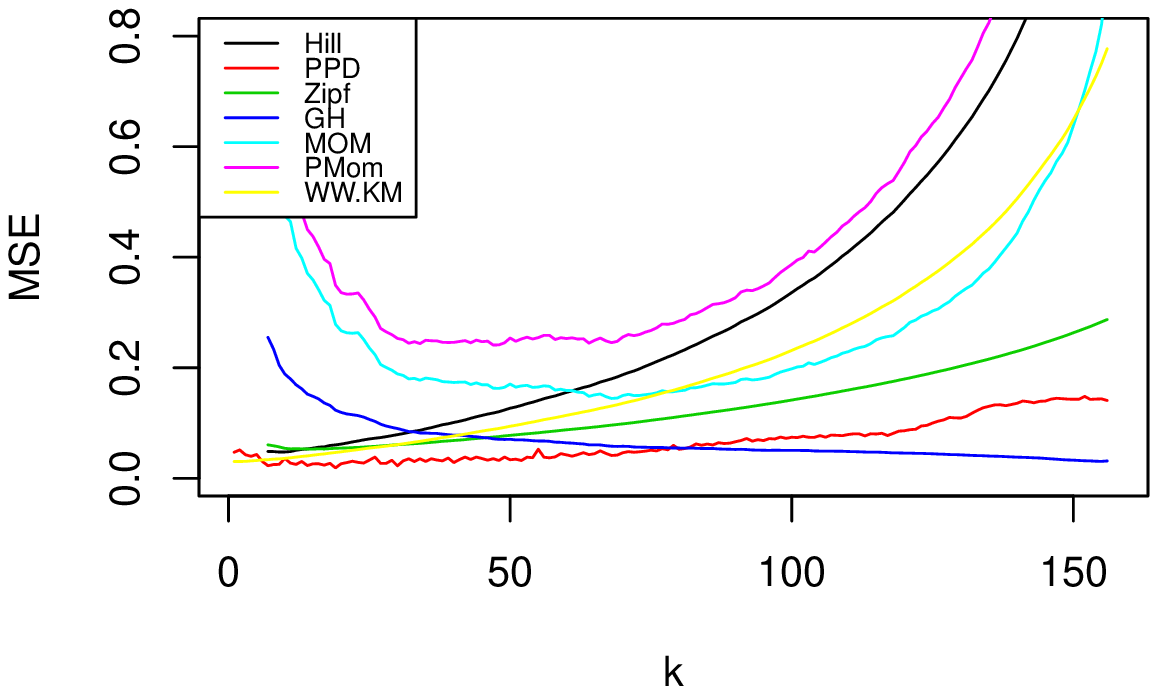}}\\
	\subfloat{%
		\includegraphics[height=4.5cm,width=.33\textwidth]{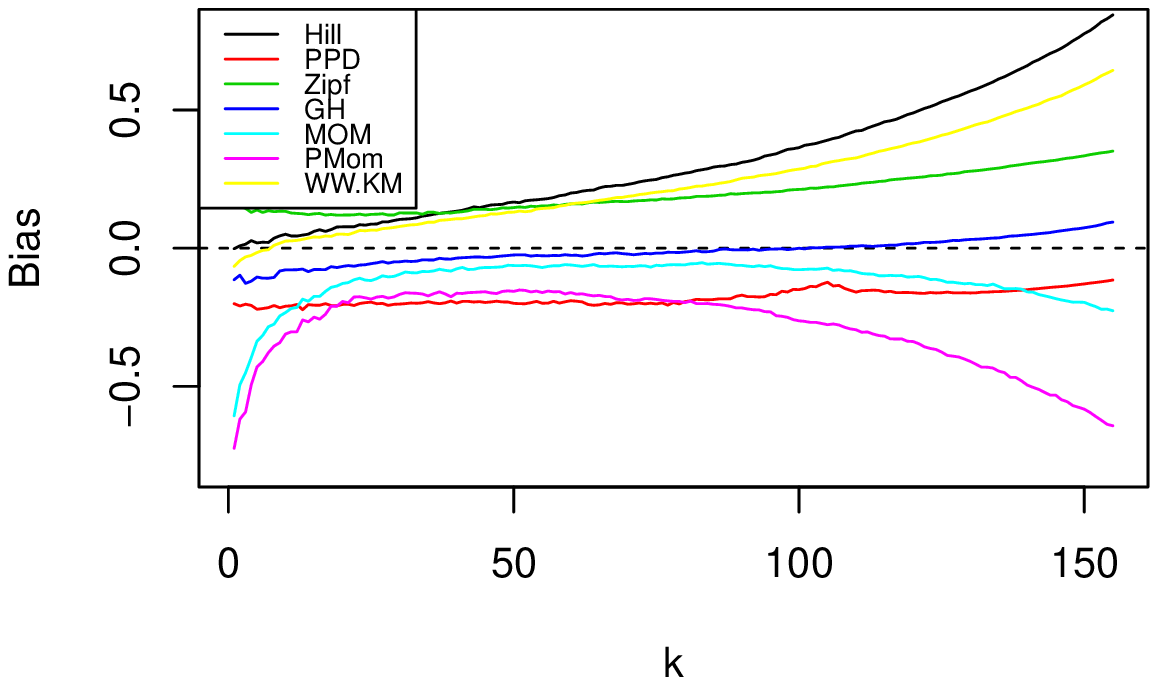}}%
	\subfloat{%
		\includegraphics[height=4.5cm,width=.33\textwidth]{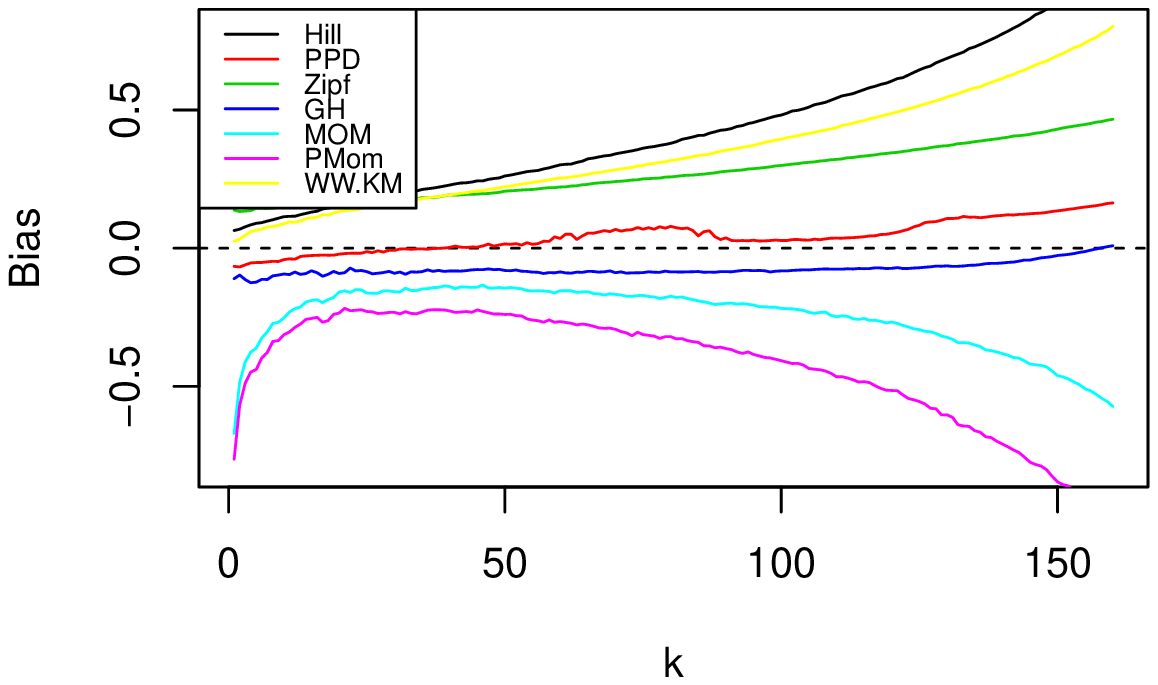}}%
	\subfloat{%
		\includegraphics[height=4.5cm,width=.33\textwidth]{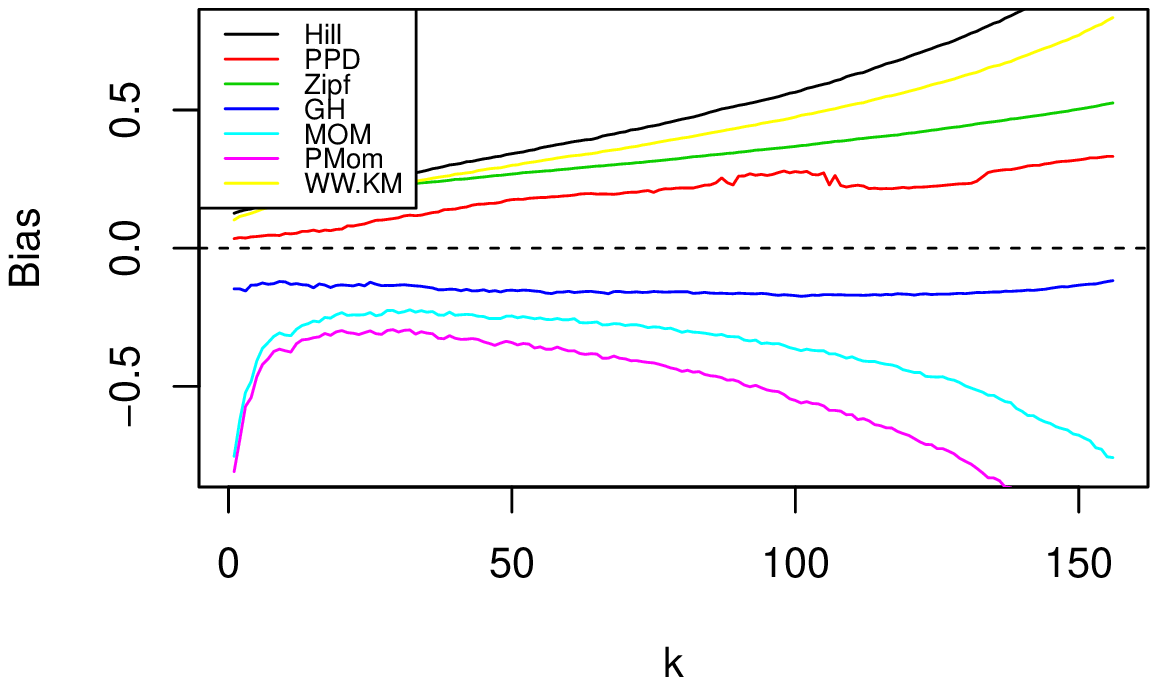}}\\
	
	\caption{ Results for Burr distribution with $\wp=0.35.$  Leftmost column: $\gamma_1(x)=0.63~ (x=0.12);$ Middlemost column : $\gamma_1(x)=0.31~ (x=0.37);$ Rightmost column: $\gamma_1(x)=0.10~ (x=0.75);$}
	\label{B2}
\end{figure}

\begin{figure}[htp!]
	\centering
	\subfloat{%
		\includegraphics[height=4.5cm,width=.33\textwidth]{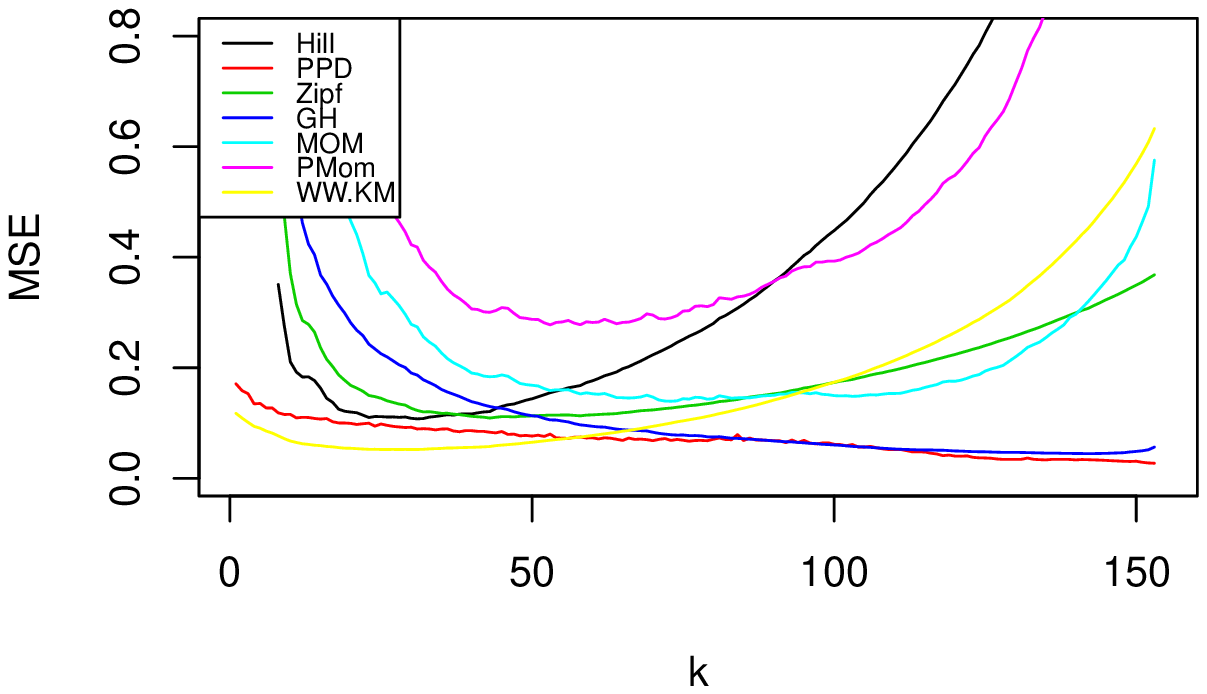}}%
	\subfloat{%
		\includegraphics[height=4.5cm,width=.33\textwidth]{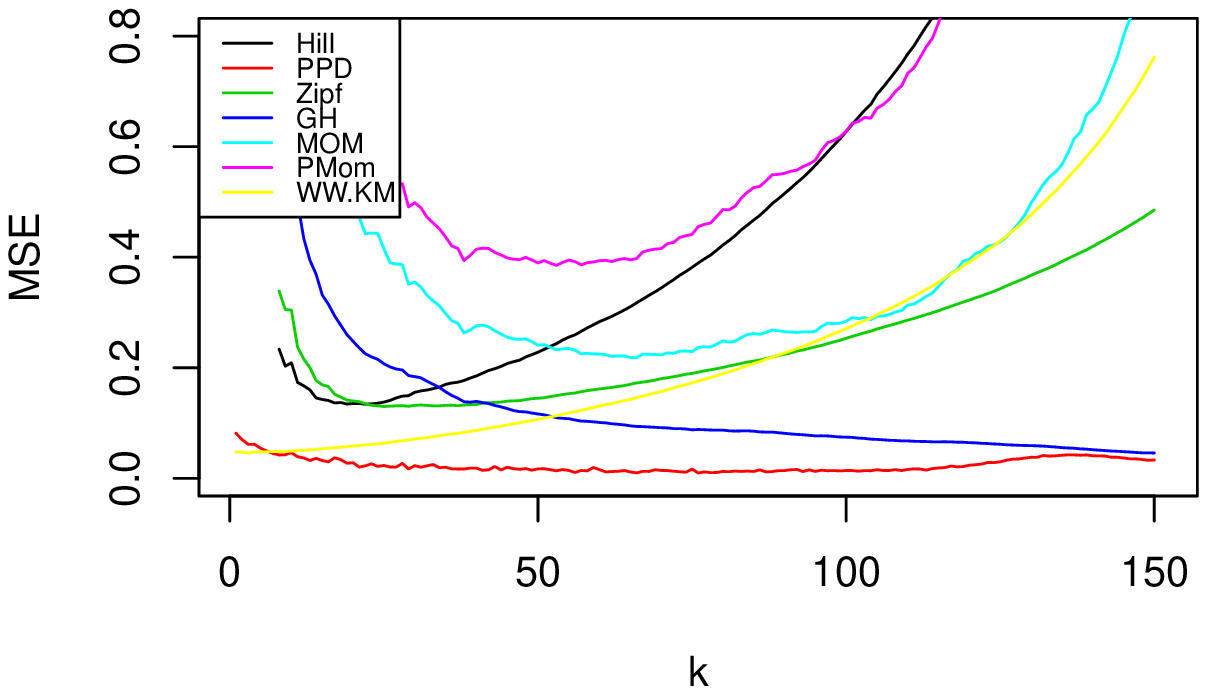}}%
	\subfloat{%
		\includegraphics[height=4.5cm,width=.33\textwidth]{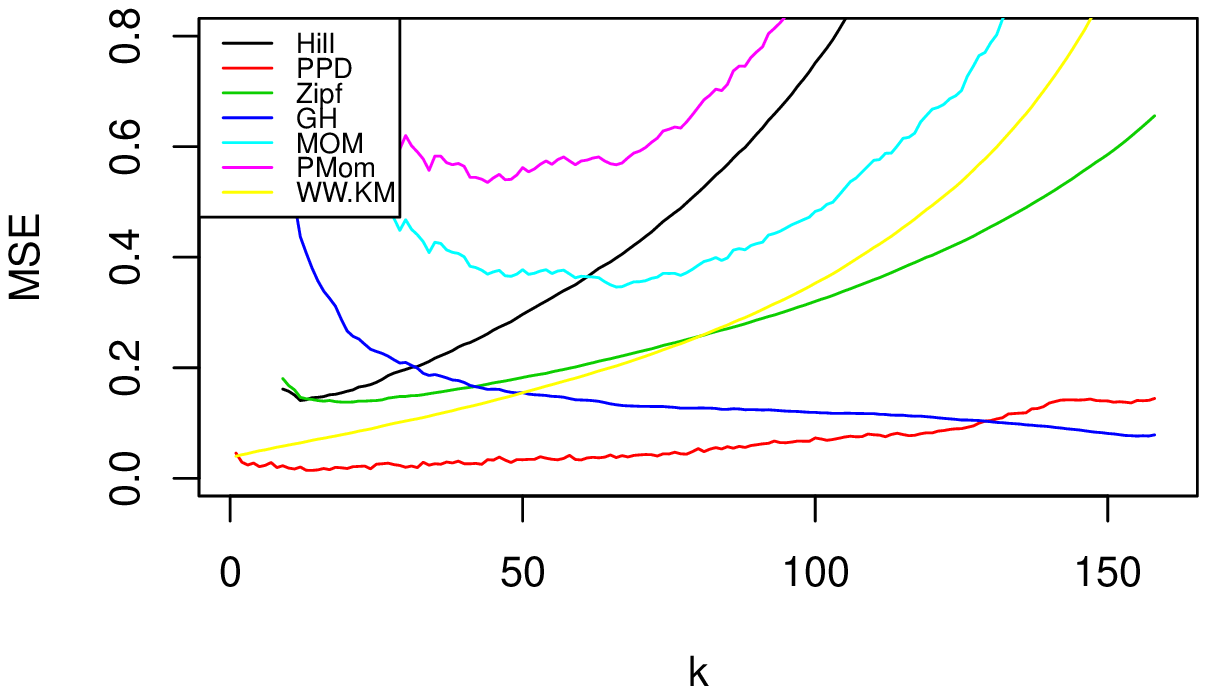}}\\
	\subfloat{%
		\includegraphics[height=4.5cm,width=.33\textwidth]{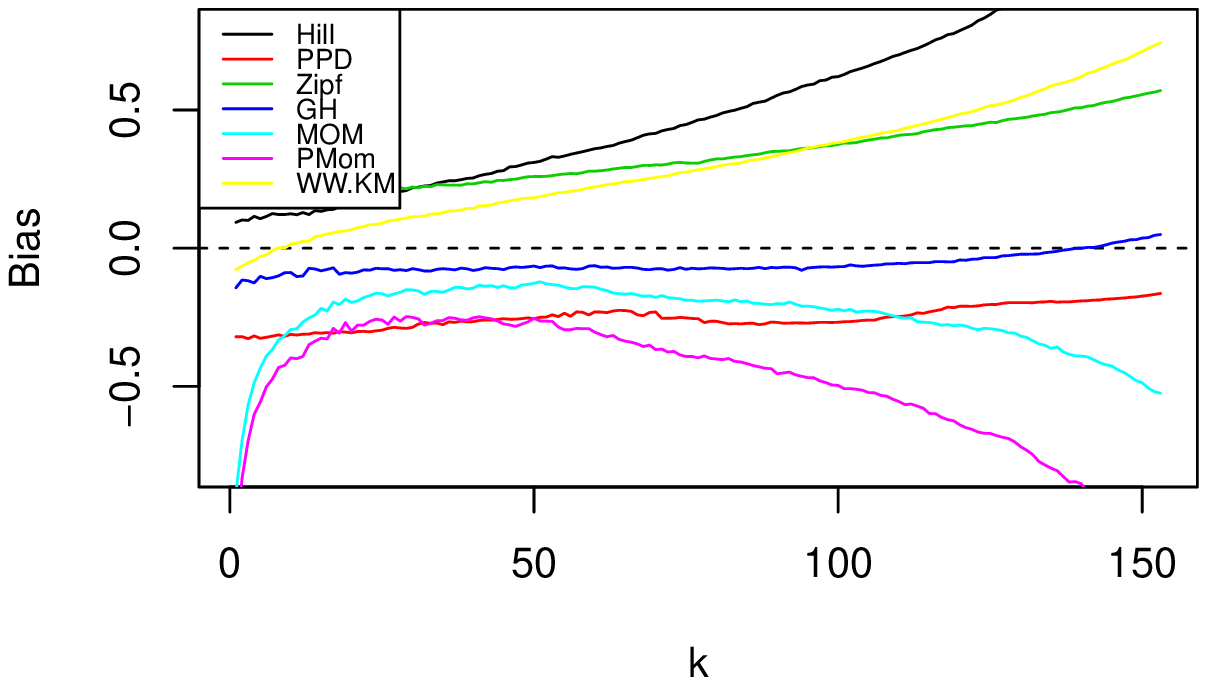}}%
	\subfloat{%
		\includegraphics[height=4.5cm,width=.33\textwidth]{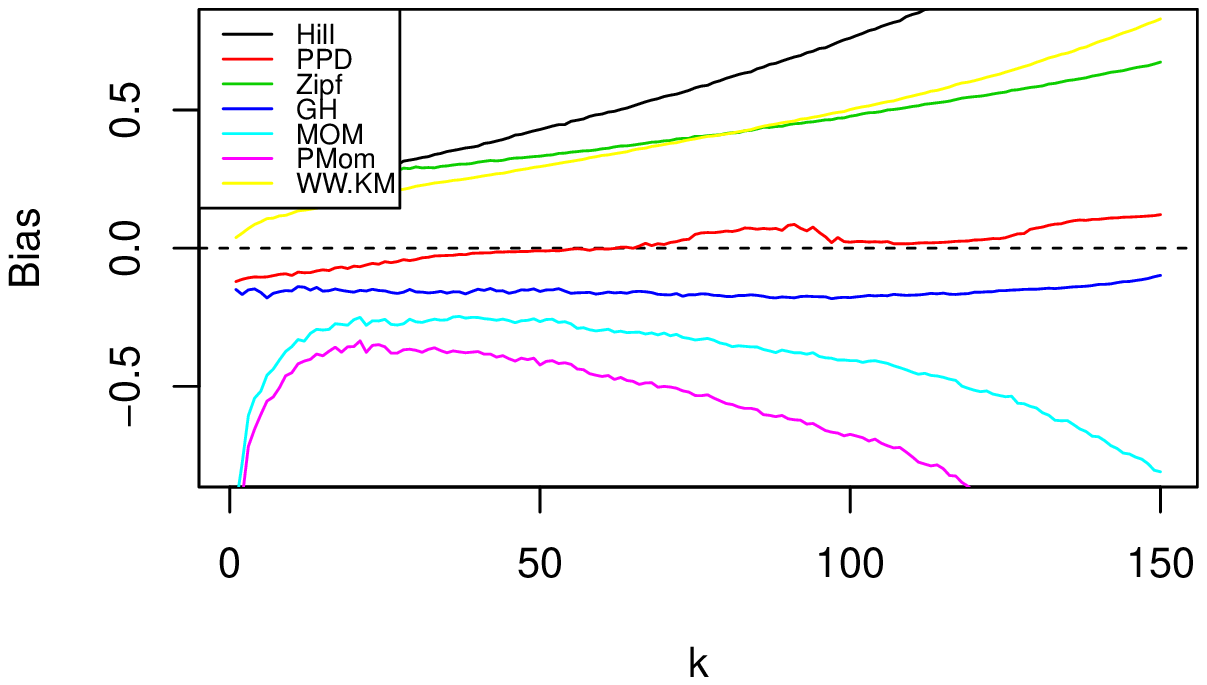}}%
	\subfloat{%
		\includegraphics[height=4.5cm,width=.33\textwidth]{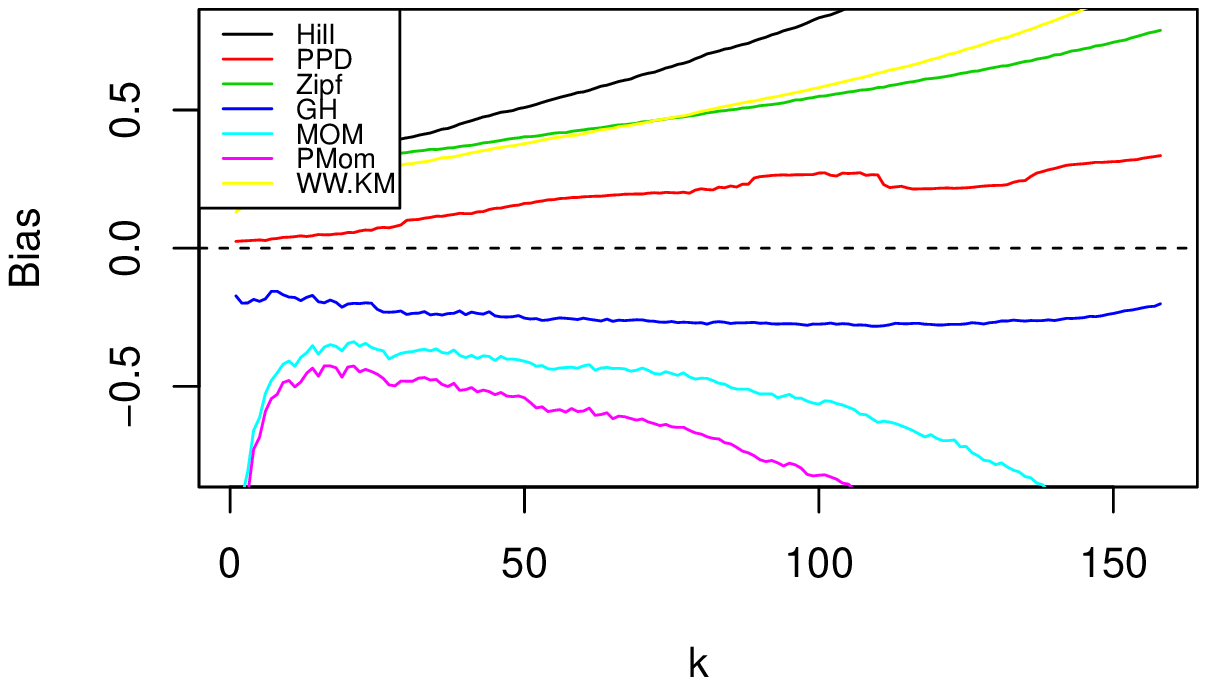}}\\
	
	\caption{ Results for Burr distribution with $\wp=0.55.$  Leftmost column: $\gamma_1(x)=0.63~ (x=0.12);$ Middlemost column : $\gamma_1(x)=0.31~ (x=0.37);$ Rightmost column: $\gamma_1(x)=0.10~ (x=0.75);$}
	\label{B3}
\end{figure}

The results of the simulation study are presented in this section. For brevity and ease of presentation, we present the results for Burr distribution and leave that of Pareto and Fr\'{e}chet to the appendix.
Figures \ref{B1}, \ref{B2} and \ref{B3} show the results for estimators of $\gamma_1(x)=0.63 ~(x=0.10),~ \gamma_1(x)=0.31~ (x=0.37)$ and $\gamma_1(x)=0.10~ (x=0.75)$ respectively. From these figures, it can easily be seen that most of the estimators' performance diminish as $k$ increases. This is expected as more intermediate observations are included in the estimation leading to bias. In addition, we observed that the bias and to a larger extent MSE, increases with decreasing value of $\gamma_1(x).$ Furthermore, the performance of the estimators of  $\gamma_1(x)$ decreases as the censoring percentage increases. This is in conformity to the findings in \citet{Ndao2014}. 

We now turn attention to the performance of the individual estimators. Firstly, we found that the Hill estimator has large MSE and bias as $k$ increases. This is in contrast to the simulation results and Corollary 4.2 in \citet{Ndao2014}. Thus, we may conclude that the performance of the Hill estimator depends on the choice of parameter function, $\gamma_1(x).$ However, this result is consistent with the performance of the Hill estimator in the case where there is no covariate information nor censoring \citep[see e.g.][and references therein]{Beirlant2004}.

For samples generated from the Burr distribution,  the PPD, GH and MOM estimators are the most robust to censoring in most cases. These estimators have smaller bias and MSE compared to the other estimators of $\gamma_1(x).$ In particular, the proposed PPD estimator is seen to have the smallest bias and MSE as the percentage of censoring increases.

With regard to the Pareto distribution, the performance of the estimators are similar to that of the Burr distribution. However, some differences occur which are mentioned. Firstly, the PMom estimator competes with the best three estimators i.e. PPD, MOM and GH in terms of bias and MSE. Secondly, in the case, of large censoring, the PPD estimator has the smallest MSE and relatively good bias.

In the case of samples generated from the Fr\'{e}chet distribution, all the estimators have good MSE values with the exception of small and large $k$ values where some of the estimators' performance deteriorate. In terms of bias, the performance does not differ significantly from that of MSE. Interestingly, the Hill estimator is seen to compete with the PPD, Zipf and WW.KM in terms of bias and MSE as the percentage of censoring and $k$ increase. 

Overall, we found that the performance of estimators depend on the distribution, the number of top order statistics and the percentage of censoring. Contrary to what was reported in \citet{Ndao2014}, we found that the Hill estimator has large bias and MSE except for samples generated from the Fr\'{e}chet distribution. The proposed PPD estimator is universally competitive in estimating $\gamma_1(x)$ regardless of its size, percentage of censoring and number of top order statistics.

\section{Conclusion}
In this paper, the central issues were a review and proposals of estimators of conditional extreme value index when observations are subject to right random censoring. In the latter, we proposed adapting some classical extreme value index to censoring and presence of covariate information. The existing and the proposed estimators were compared in a simulation study. Some interesting results were obtained and are outlined as follows:

\begin{enumerate}

\item The performance of the estimators depend on the underlying distribution of the sampled data. Thus, no estimator was universally the best under all the simulation conditions considered. However, a closer look at the results reveal that some general conclusions can be reached on some estimators that can be considered as appropriate for estimating $\gamma_1(x).$ 
	
 \item The performance of the estimators depend on the size of $\gamma_1(x):$ Bias and MSE values were generally larger for small $\gamma_1(x)$ values and smaller for larger values of $\gamma_1(x).$ Therefore, we recommend that in practice before proceeding to use any estimator for $\gamma_1(x),$ one should assess the potential range of the true value of $\gamma_1(x).$ In such cases, several estimators can be considered and a median or average value used as an estimate to help in the selection of the preferred estimator. 
 
 \item The performance of the estimators generally deteriorates as the percentage of censoring increases. However, the estimators that are robust to censoring, to a larger extent, maintain their performance under increased percentage of censoring. These estimators include PPD, MOM and PMom and GH. Conspicuously missing is the Hill estimator which was shown to have large bias and MSE except for samples from the Fr\'{e}chet distribution.
 
 \item Overall, we found that the proposed PPD estimator mostly has the best MSE and bias behaviour under small and heavy censoring. In addition, it has less bias as the number of top order statistics, $k,$ increases.

\end{enumerate}
Some additional research is needed to establish the consistency and asymptotic normality of the PPD estimator. This will enhance statistical inference and will be considered in future research.

\newpage	
\bibliographystyle{apalike}
\bibliography{CEN}

\begin{thebibliography}{}

\bibitem[Ameraoui et~al., 2016]{Ameraoui2016}
Ameraoui, A., Boukhetala, K., and Dupuy, J.-F. (2016).
\newblock Bayesian estimation of the tail index of a heavy tailed distribution
  under random censoring.
\newblock {\em Computational Statistics \& Data Analysis}, 104:148--168.

\bibitem[Beirlant et~al., 1999]{Beirlant1999a}
Beirlant, J., Dierckx, G., Goegebeur, Y., and Matthys, G. (1999).
\newblock {Tail index estimation and an exponential regression model}.
\newblock {\em Extremes}, 2:177--200.

\bibitem[Beirlant and Goegebeur, 2003]{Beirlant2003}
Beirlant, J. and Goegebeur, Y. (2003).
\newblock {Regression with response distributions of Pareto-type}.
\newblock {\em Computational Statistics and Data Analysis}, 42:595--619.

\bibitem[Beirlant and Goegebeur, 2004]{Beirlant2004b}
Beirlant, J. and Goegebeur, Y. (2004).
\newblock {Local polynomial maximum likelihood estimation for Pareto-type
  distributions}.
\newblock {\em Journal of Multivariate Analysis}, 89(1):97--118.

\bibitem[Beirlant et~al., 2004]{Beirlant2004}
Beirlant, J., Goegebeur, Y., Segers, J., and Teugels, J.~L. (2004).
\newblock {\em {Statistics of Extremes: Theory and Applications}}.
\newblock Wiley, England.

\bibitem[Beirlant et~al., 2007]{Beirlant2007}
Beirlant, J., Guillou, A., Dierckx, G., and Fils-Villetard, A. (2007).
\newblock {Estimation of the extreme value index and extreme quantiles under
  random censoring}.
\newblock {\em Extremes}, 10:151--174.

\bibitem[Beirlant et~al., 2010]{Beirlant2010}
Beirlant, J., Guillou, A., and Toulemonde, G. (2010).
\newblock {Peaks-Over-Threshold modeling under random censoring}.
\newblock {\em Communications in Statistics - Theory and Methods},
  39(7):1158--1179.

\bibitem[Beirlant et~al., 2017]{Beirlant2017}
Beirlant, J., Maribe, G., and Verster, A. (2017).
\newblock {Penalized bias reduction in extreme value estimation for censored
  Pareto-type data , and long-tailed insurance applications}.
\newblock {\em arXiv1705.0663v1}, pages 1--21.

\bibitem[Beirlant et~al., 1996]{Beirlant1996}
Beirlant, J., Vynckier, P., and Teugels, J.~L. (1996).
\newblock {Excess functions and estimation of the extreme-value index}.
\newblock {\em Bernoulli}, 2(4):293--318.

\bibitem[Brahimi et~al., 2013]{Brahimi2013}
Brahimi, B., Meraghni, D., and Necir, A. (2013).
\newblock {On the asymptotic normality of Hill ' s estimator of the tail index
  under random censoring}.
\newblock {\em arXiv1302.1666v1}, pages 1--11.

\bibitem[Danielsson et~al., 1996]{Danielsson1996}
Danielsson, J., Jansen, D.~W., and de~Vries, C.~G. (1996).
\newblock {The method of moments ratio estimator for the tail shape parameter}.
\newblock {\em Communications in Statistics - Theory and Methods},
  25(4):711--720.

\bibitem[Davison and Smith, 1990]{Davison1990}
Davison, A.~C. and Smith, R.~L. (1990).
\newblock {Models for exceedances over high thresholds}.
\newblock {\em Journal of the Royal Statistical Society: Series B (Statistical
  Methodology)}, 52(3):393--442.

\bibitem[de~Haan, 1970]{deHaan1970}
de~Haan, L. (1970).
\newblock {\em {On regular variation and its application to the weak
  convergence of sample extremes}}.
\newblock Phd, University of Amsterdam.

\bibitem[de~Haan and Peng, 1998]{deHaan1998}
de~Haan, L. and Peng, L. (1998).
\newblock {Comparison of tail index estimators}.
\newblock {\em Statistica Neerlandica}, 52(1):60--70.

\bibitem[Deheuvels et~al., 1997]{Deheuvels1997}
Deheuvels, P., de~Haan, L., Peng, L., and Pereira, T.~T. (1997).
\newblock {Comparison of extreme value index estimators}.
\newblock Technical Report T400:EUR-09, The Erasmus University Rotterdam,
  Rotterdam.

\bibitem[Dekkers et~al., 1989]{Dekkers1989}
Dekkers, A. L.~M., Einmahl, J. H.~J., and de~Haan, L. (1989).
\newblock {A moment estimator for the index of an extreme-value distribution}.
\newblock {\em Annals of Statistics}, 17(4):1833--1855.

\bibitem[Einmahl et~al., 2008]{Einmahl2008}
Einmahl, J. H.~J., Elie, A.~M., and Guillou, A. (2008).
\newblock {Statistics of extremes under random censoring}.
\newblock {\em Bernoulli}, 14(1):207--227.

\bibitem[Fisher and Tippett, 1928]{Fisher1928}
Fisher, R. and Tippett, L. (1928).
\newblock {On the estimation of the frequency distributions of the largest or
  smallest member of a sample}.
\newblock {\em Proceedings of the Cambridge Philosophical Society}, 24:80--190.

\bibitem[Gardes and Girard, 2008]{Gardes2008}
Gardes, L. and Girard, S. (2008).
\newblock {A moving window approach for nonparametric estimation of the
  conditional tail index}.
\newblock {\em Journal of Multivariate Analysis}, 99(10):2368--2388.

\bibitem[Gilli and K{\"{e}}llezi, 2006]{Gilli2006}
Gilli, M. and K{\"{e}}llezi, E. (2006).
\newblock {An application of extreme value theory for measuring financial
  risk}.
\newblock {\em Computational Economics}, 27(1):1--23.

\bibitem[Gnedenko, 1943]{Gnedenko1943}
Gnedenko, B. (1943).
\newblock {Sur la distribution limite du terme Maximum d'une s\'{e}rie
  al\'{e}atoire}.
\newblock {\em Annals of Mathematics}, 44(3):423--453.

\bibitem[Gomes and Guillou, 2014]{Gomes2014}
Gomes, M.~I. and Guillou, A. (2014).
\newblock {Extreme value theory and statistics of univariate extremes : a
  review}.
\newblock {\em International Statistical Review}, pages 1--35.

\bibitem[Gomes et~al., 2008]{Gomes2008}
Gomes, M.~I., Lu{\'{i}}sa, C. e.~C., {Fraga Alves}, M.~I., and Pestana, D.
  (2008).
\newblock {Statistics of extremes for IID data and breakthroughs in the
  estimation of the extreme value index: Laurens de Haan leading
  contributions}.
\newblock {\em Extremes}, 11(1):3--34.

\bibitem[Gomes and Neves, 2011]{Gomes2011}
Gomes, M.~I. and Neves, C. (2011).
\newblock {Estimation of the extreme value index for randomly censored data}.
\newblock {\em Biometrical Letters}, 48(1):1--22.

\bibitem[Hill, 1975]{Hill1975}
Hill, B. (1975).
\newblock {A simple general approach to inference about the tail of a
  distribution}.
\newblock {\em Annals of Statistics}, 3:1163--1174.

\bibitem[Kratz and Resnick, 1996]{Kratz1996}
Kratz, M. and Resnick, S.~I. (1996).
\newblock {The qq-estimator of the index of regular variation}.
\newblock {\em Communications in Statistics: Stochastic Models}, 12:699--724.

\bibitem[Ndao et~al., 2014]{Ndao2014}
Ndao, P., Diop, A., and Dupuy, J.~F. (2014).
\newblock {Nonparametric estimation of the conditional tail index and extreme
  quantiles under random censoring}.
\newblock {\em Computational Statistics and Data Analysis}, 79:63--79.

\bibitem[Ndao et~al., 2016]{Ndao2016}
Ndao, P., Diop, A., and Dupuy, J.-f. (2016).
\newblock {Nonparametric estimation of the conditional extreme-value index with
  random covariates and censoring}.
\newblock {\em Journal of Statistical Planning and Inference}, 168:20--37.

\bibitem[Stupfler, 2016]{Stupfler2016}
Stupfler, G. (2016).
\newblock {Estimating the conditional extreme-value index under random
  right-censoring}.
\newblock {\em Journal of Multivariate Analysis}, 144:1--24.

\bibitem[Tsourti and Panaretos, 2003]{Panaretos2003}
Tsourti, Z. and Panaretos, J. (2003).
\newblock {Extreme Value Index Estimators and Smoothing Alternatives: Review
  and Simulation Comparison}.
\newblock In Panaretos, J., editor, {\em Stochastic Musings: Perspectives From
  the Pioneers of the Late 20th Century}, pages 141--160. Mahwah, New Jersey.

\bibitem[Wang and Tsai, 2009]{Wang2009}
Wang, H. and Tsai, C.-L. (2009).
\newblock {Tail index regression}.
\newblock {\em Journal of the American Statistical Association},
  104(487):1233--1240.

\bibitem[Worms and Worms, 2014]{Worms2014}
Worms, J. and Worms, R. (2014).
\newblock {New estimators of the extreme value index under random right
  censoring, for heavy-tailed distributions}.
\newblock {\em Extremes}, 17(2):337--358.

\end{thebibliography}

\begin{appendices}
	\addtocontents{toc}{\protect\setcounter{tocdepth}{0}}

\section{Pareto Distribution}

For each figure, the following description of the panels apply.  Leftmost column: $\gamma_1(x)=0.63~ (x=0.12);$ Middlemost column : $\gamma_1(x)=0.31~ (x=0.37);$ Rightmost column: $\gamma_1(x)=0.10~ (x=0.75);$	
\begin{figure}[H]
	\centering
	\subfloat{%
		\includegraphics[height=4.5cm,width=.33\textwidth]{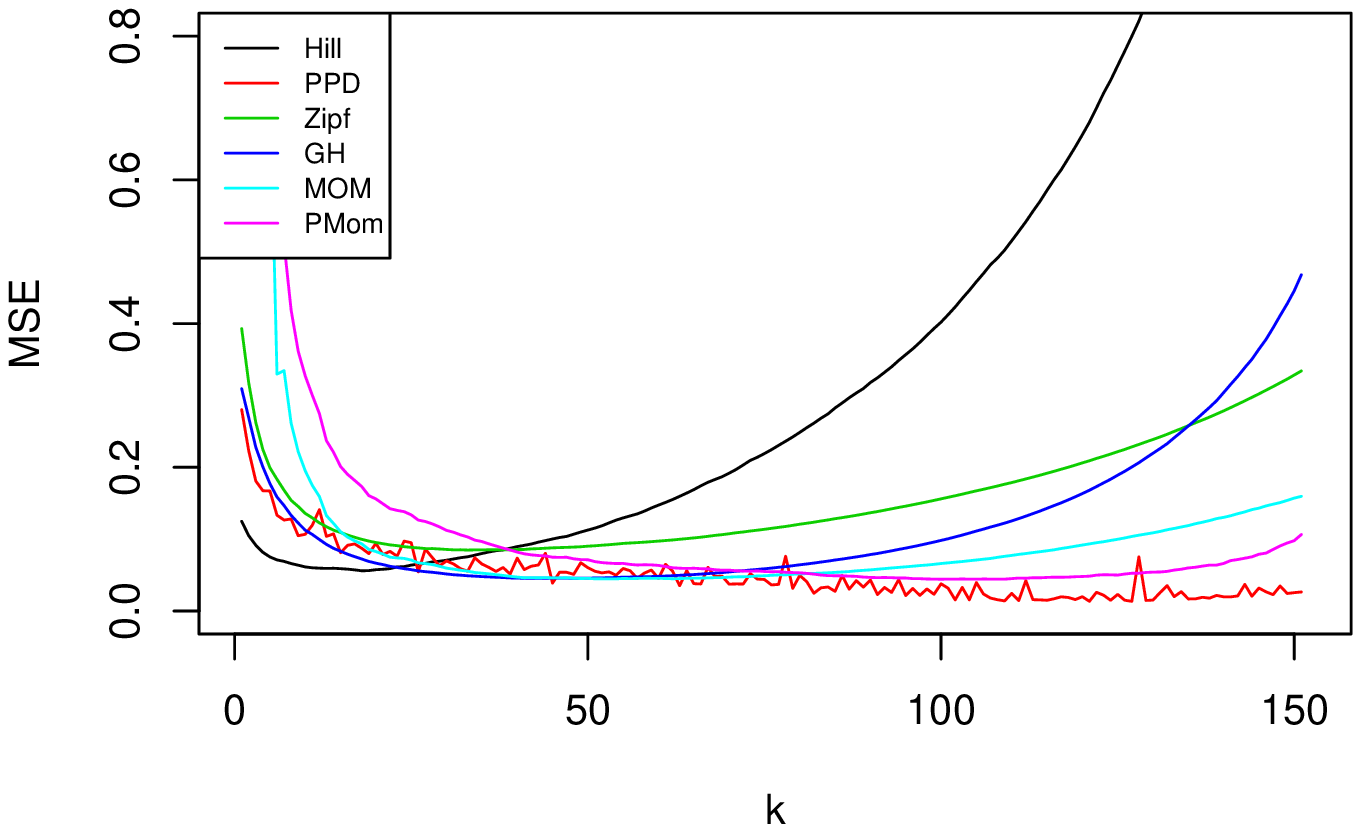}}%
	\subfloat{%
		\includegraphics[height=4.5cm,width=.33\textwidth]{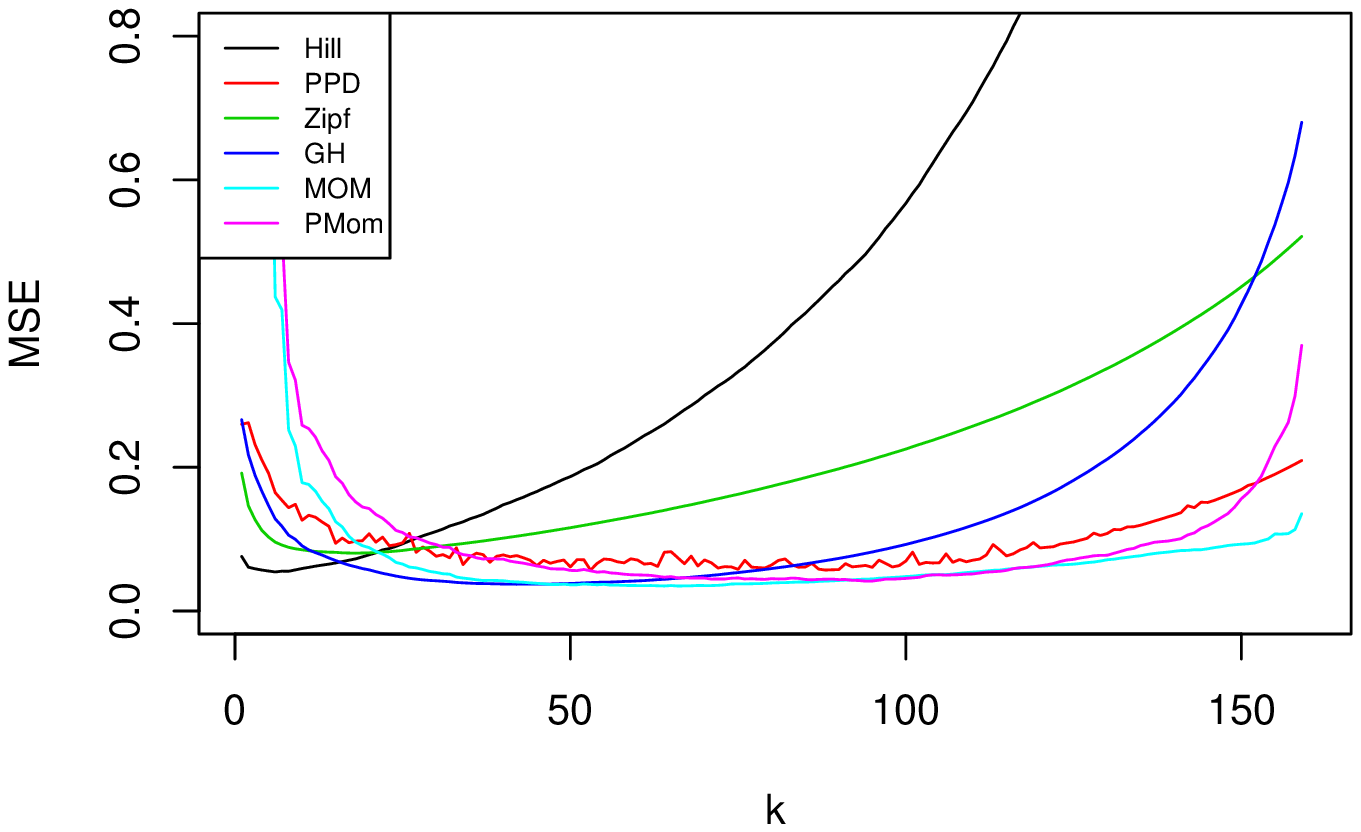}}%
	\subfloat{%
		\includegraphics[height=4.5cm,width=.33\textwidth]{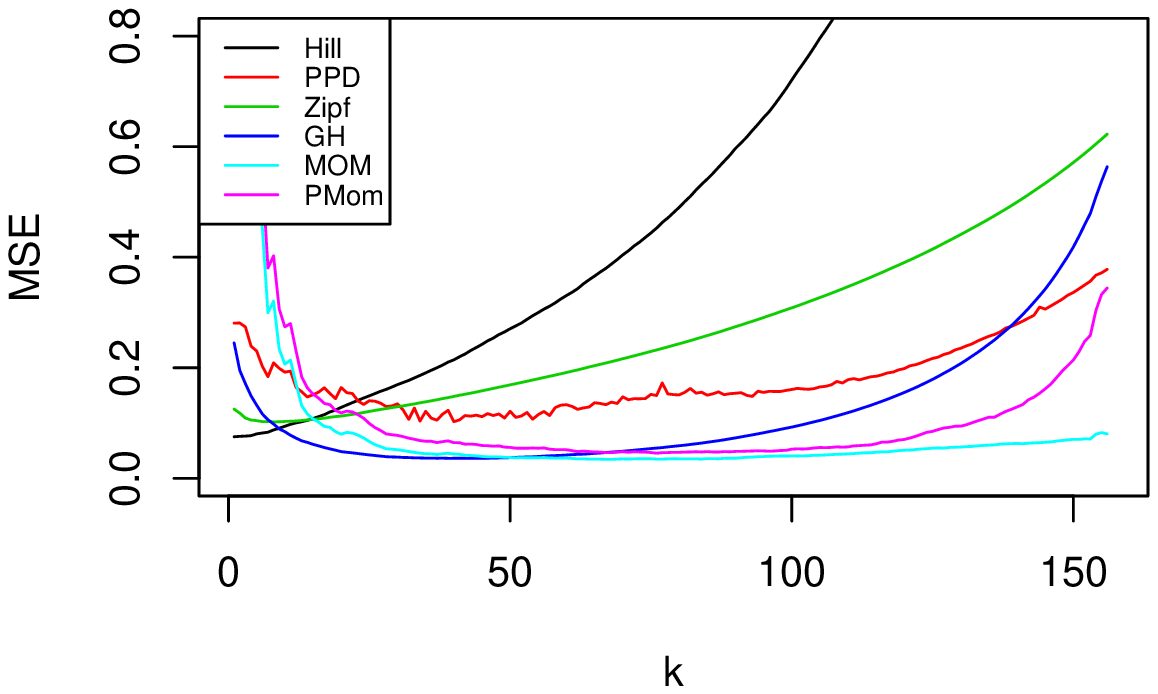}}\\
	\subfloat{%
		\includegraphics[height=4.5cm,width=.33\textwidth]{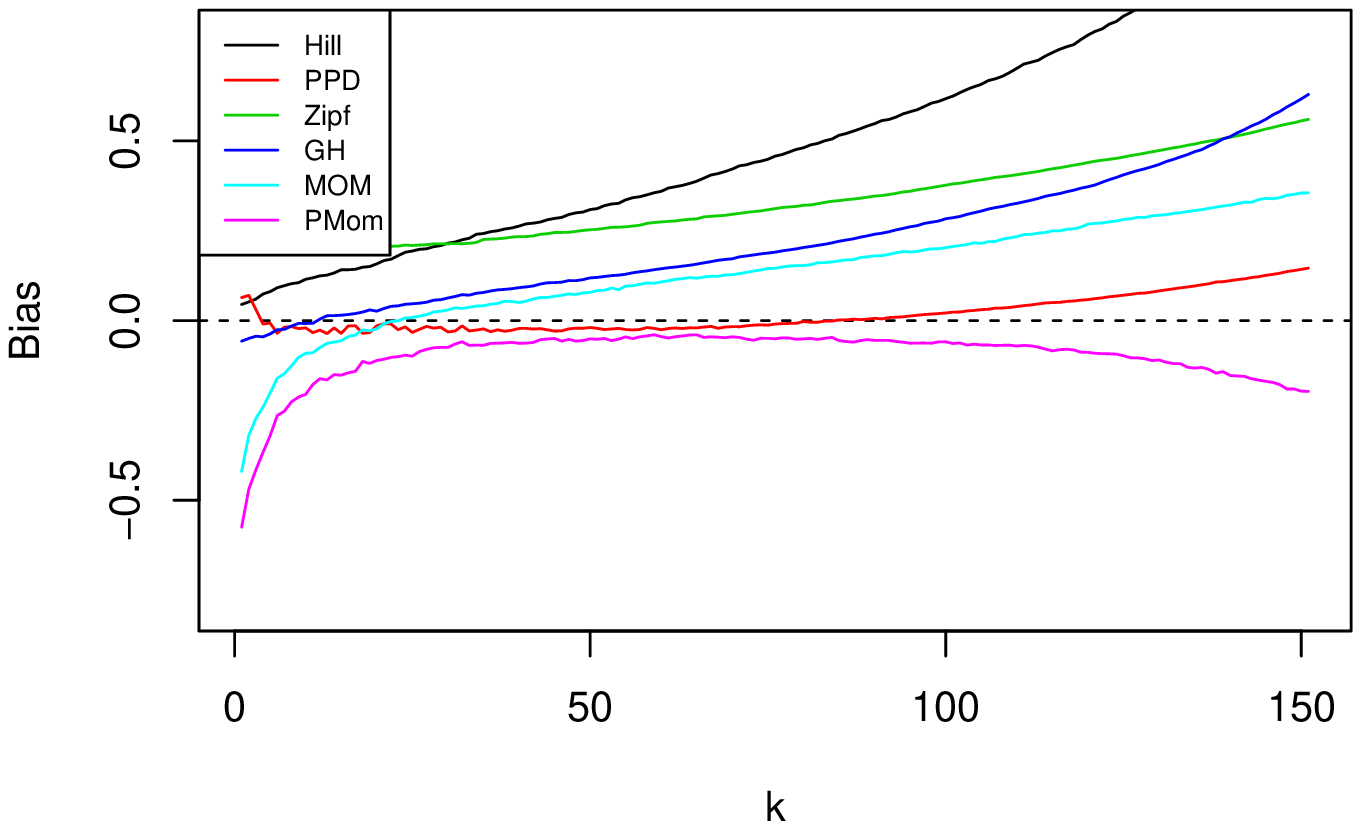}}%
	\subfloat{%
		\includegraphics[height=4.5cm,width=.33\textwidth]{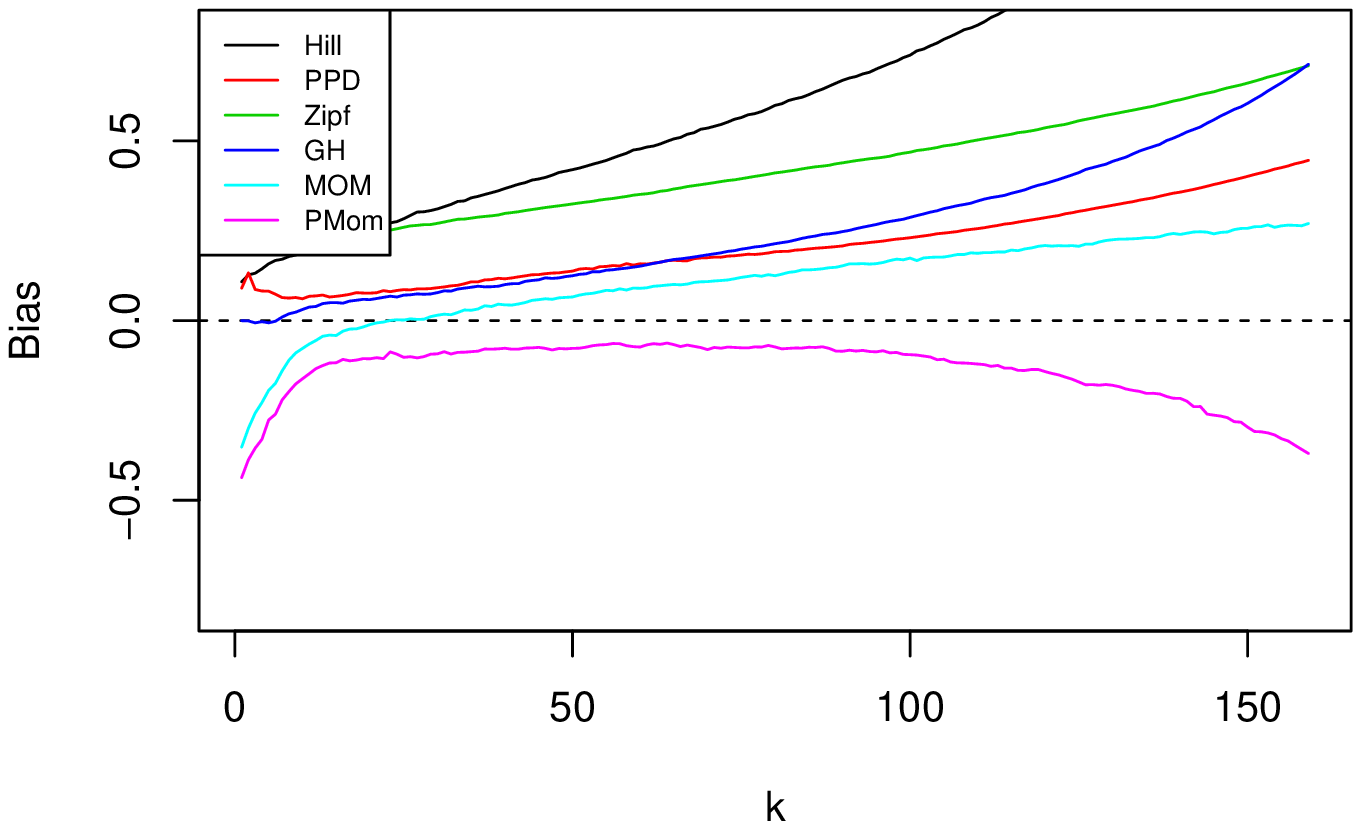}}%
	\subfloat{%
		\includegraphics[height=4.5cm,width=.33\textwidth]{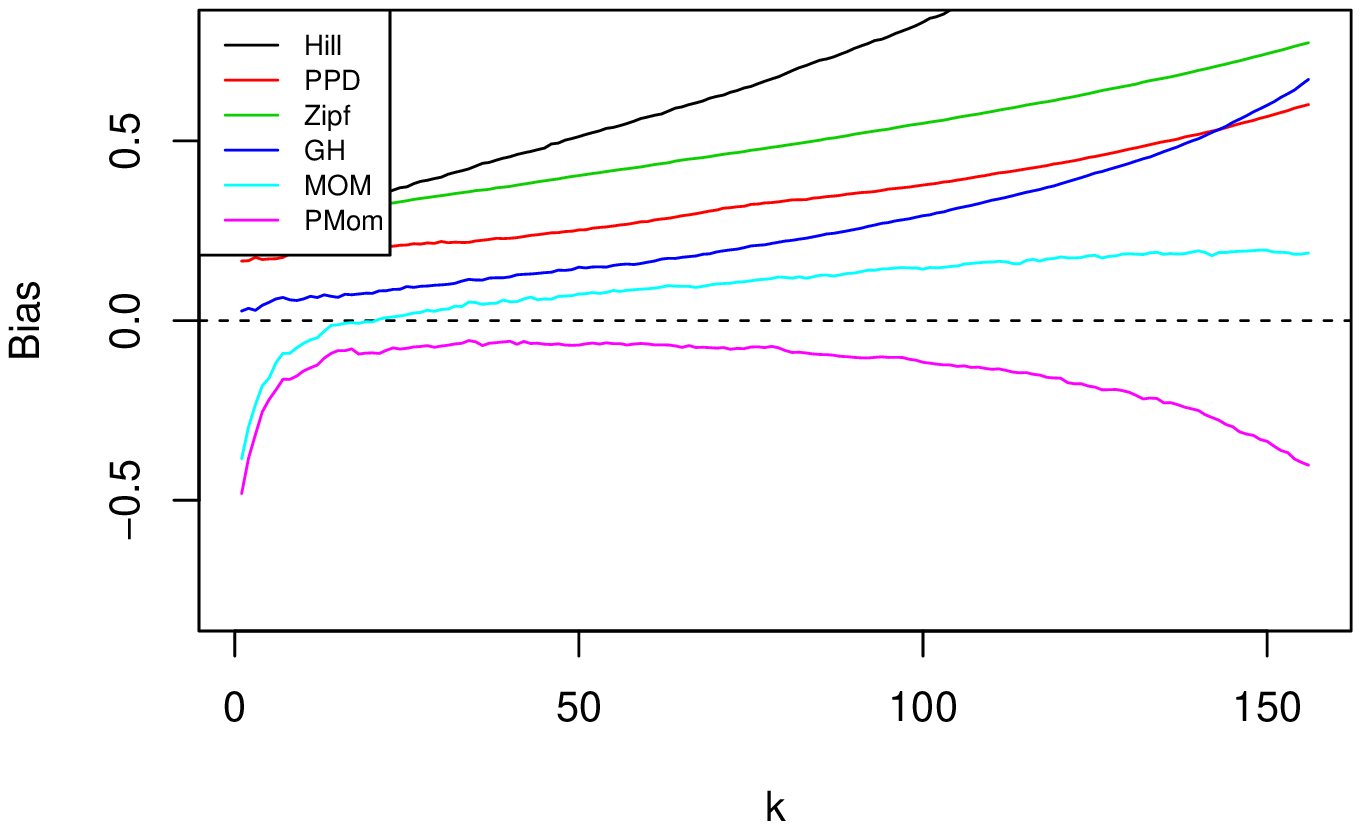}}\\
	
	\caption{ Results for Pareto distribution with $\wp=0.1.$ }
	\label{}
\end{figure}


\begin{figure}[H]
	\centering
	\subfloat{%
		\includegraphics[height=4.5cm,width=.33\textwidth]{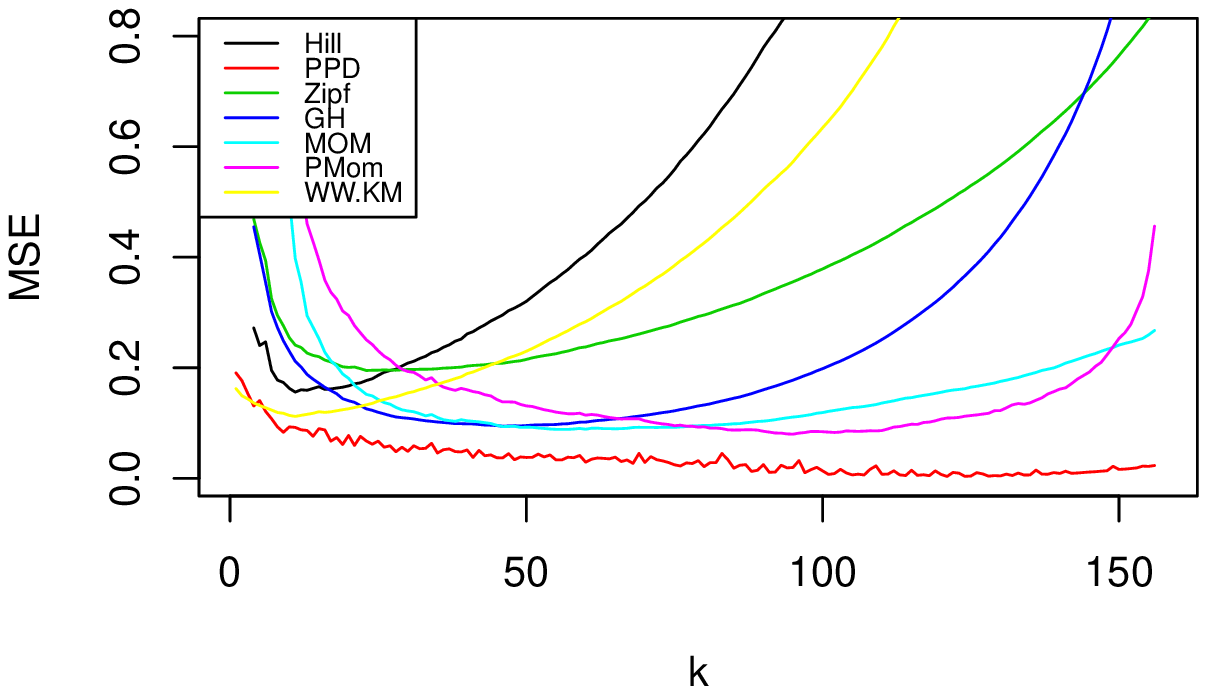}}%
	\subfloat{%
		\includegraphics[height=4.5cm,width=.33\textwidth]{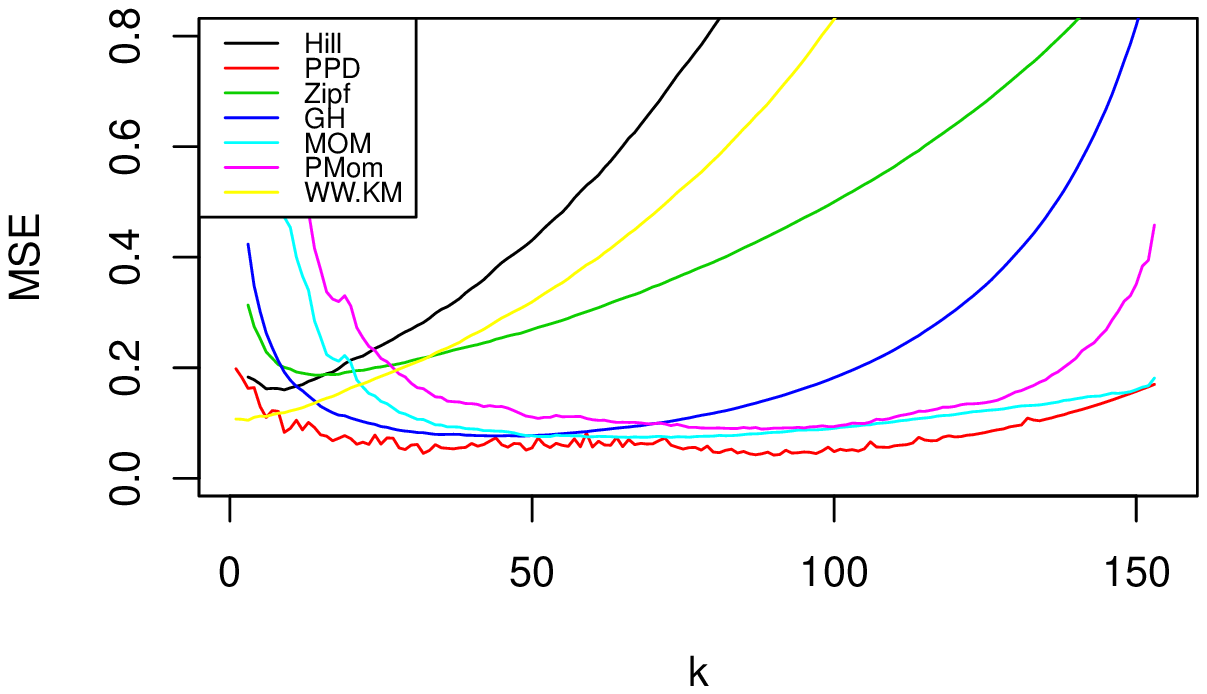}}%
	\subfloat{%
		\includegraphics[height=4.5cm,width=.33\textwidth]{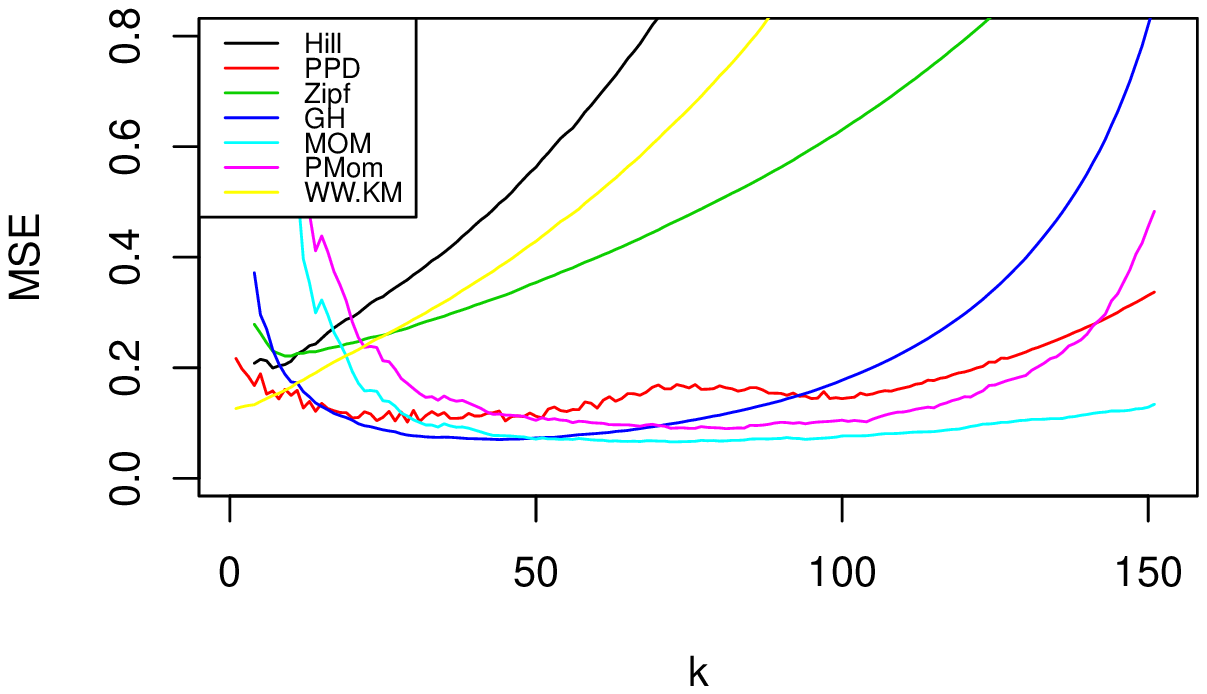}}\\
	\subfloat{%
		\includegraphics[height=4.5cm,width=.33\textwidth]{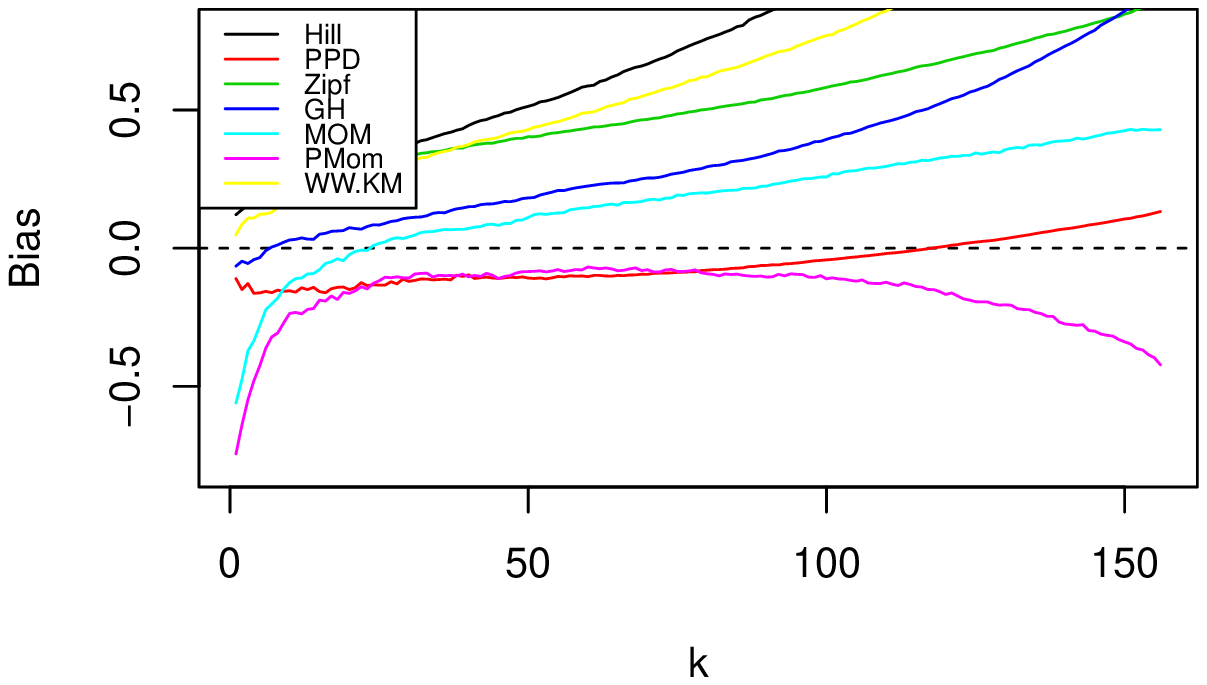}}%
	\subfloat{%
		\includegraphics[height=4.5cm,width=.33\textwidth]{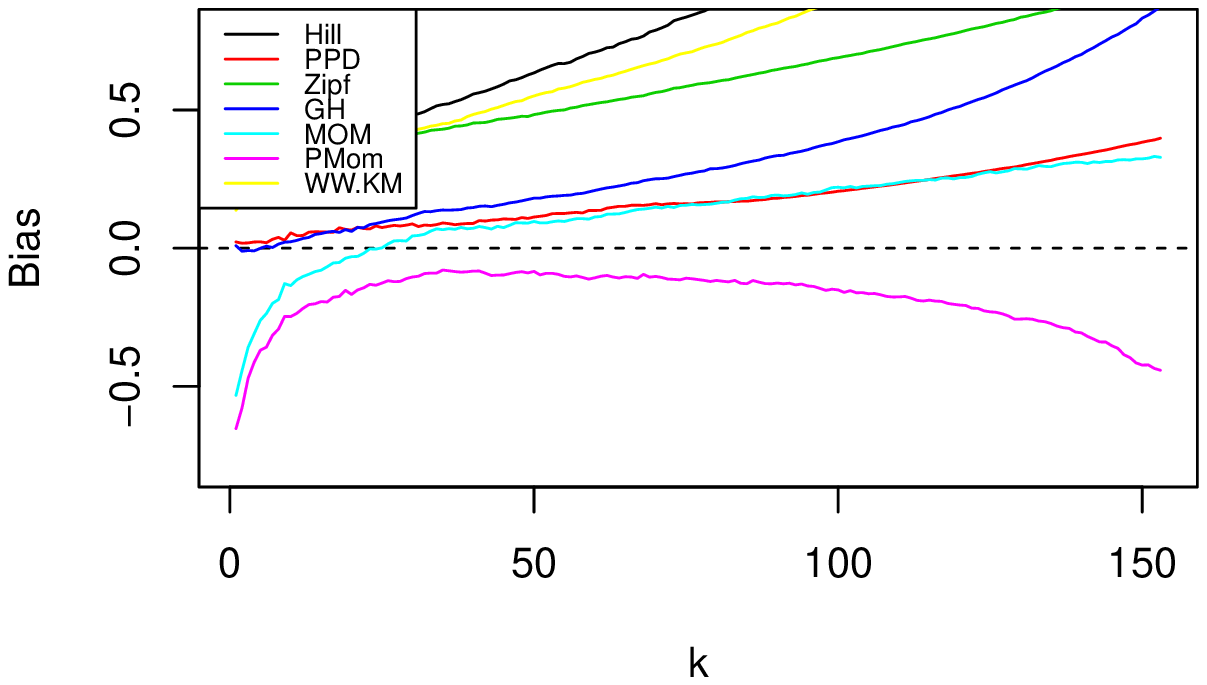}}%
	\subfloat{%
		\includegraphics[height=4.5cm,width=.33\textwidth]{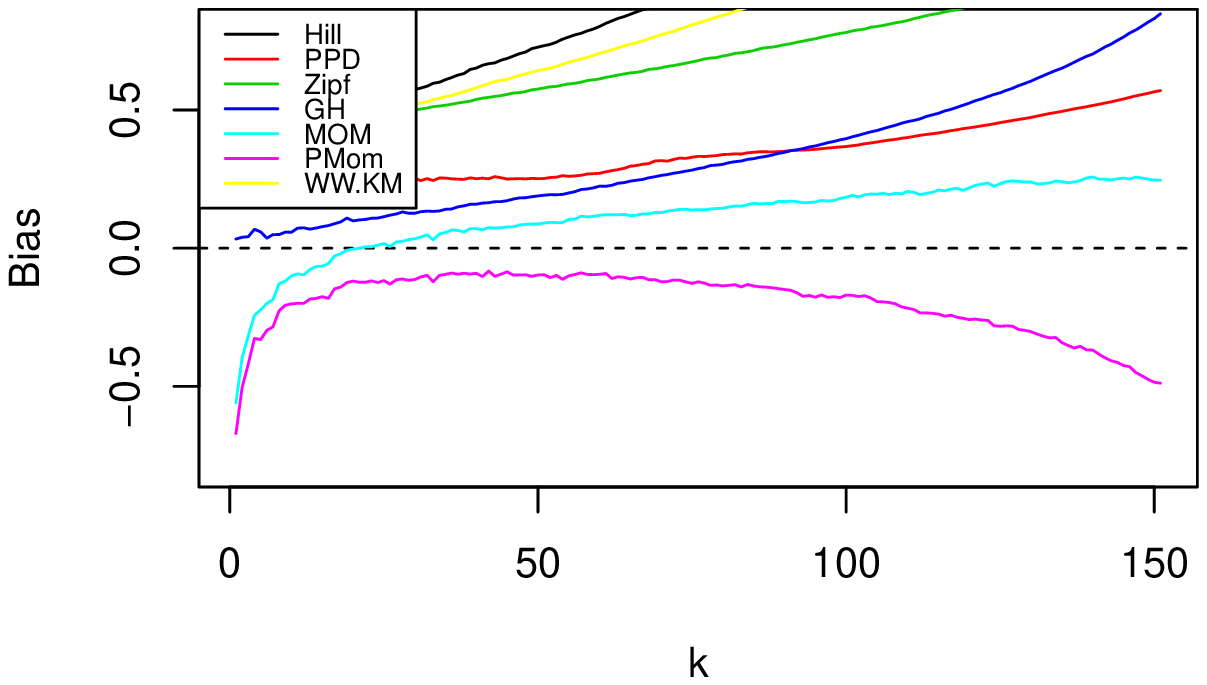}}\\
	
	\caption{ Results for Pareto distribution with $\wp=0.35.$ }
	\label{}
\end{figure}


\begin{figure}[H]
	\centering
	\subfloat{%
		\includegraphics[height=4.5cm,width=.33\textwidth]{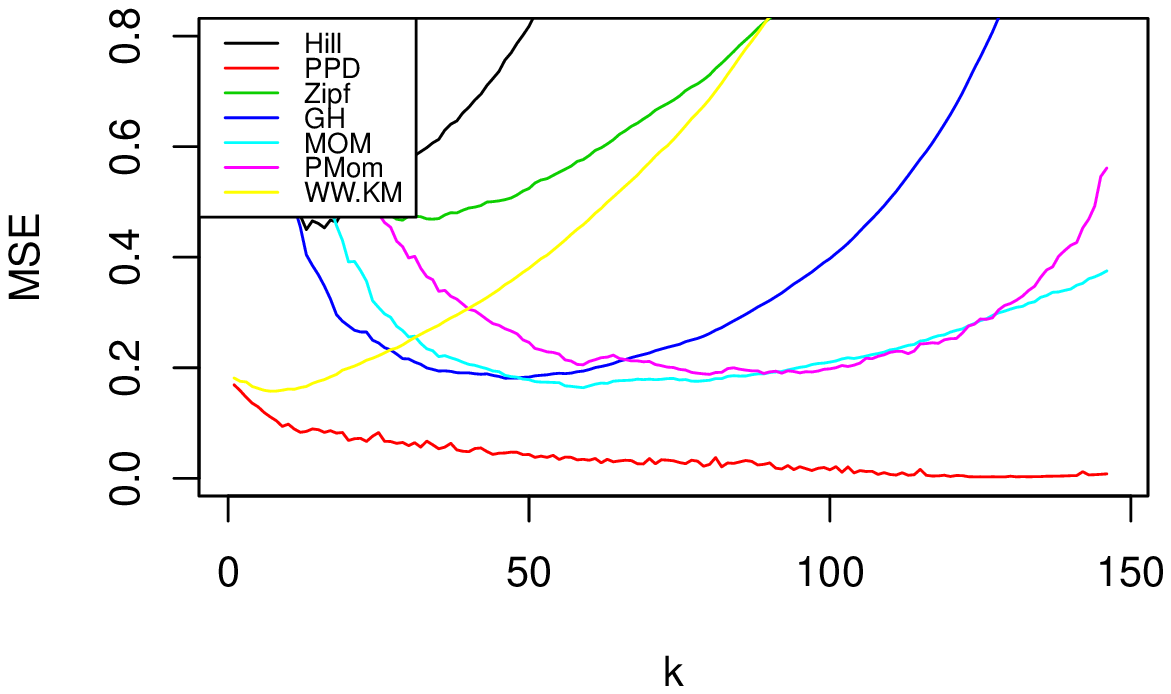}}%
	\subfloat{%
		\includegraphics[height=4.5cm,width=.33\textwidth]{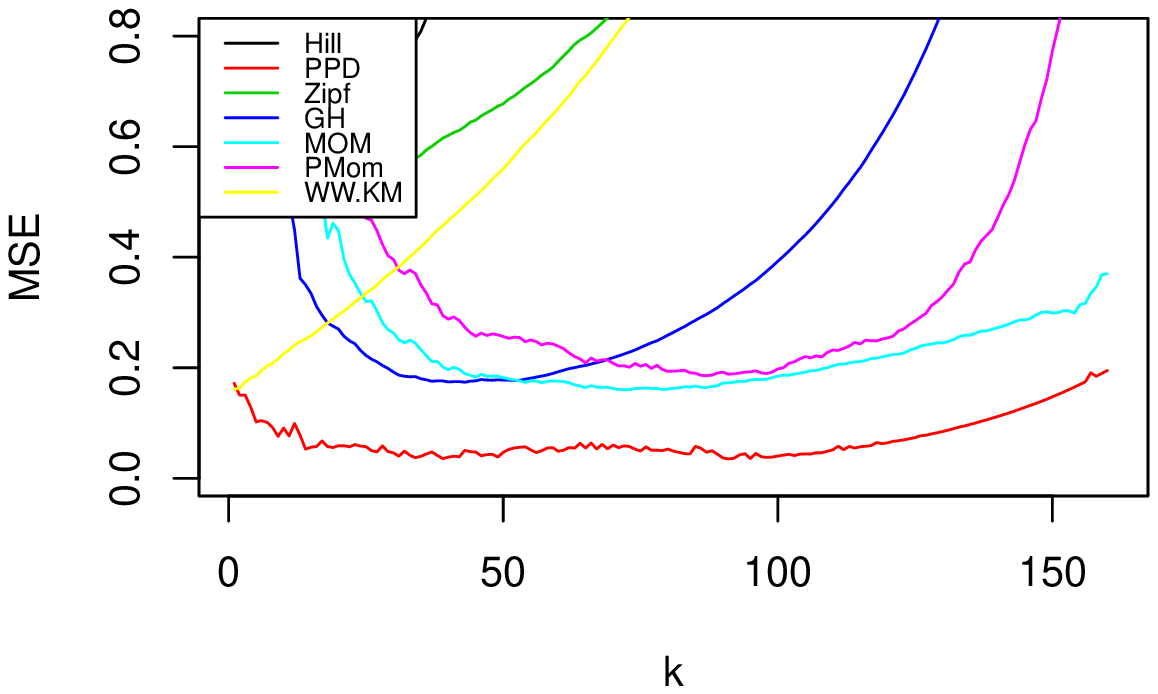}}%
	\subfloat{%
		\includegraphics[height=4.5cm,width=.33\textwidth]{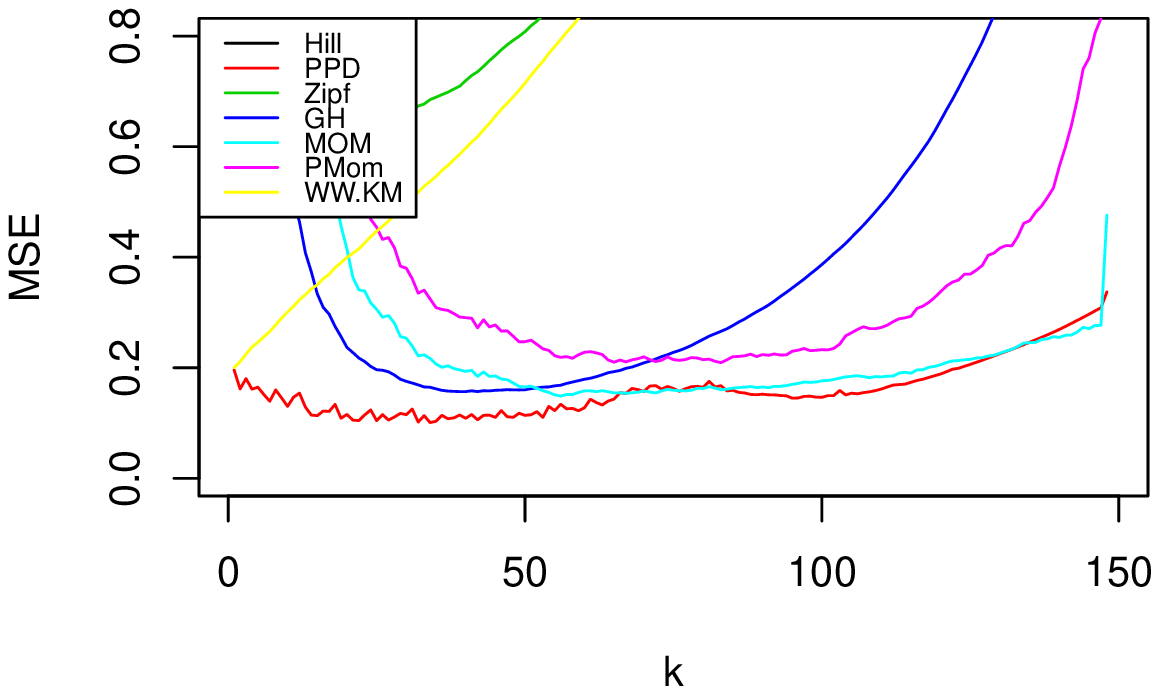}}\\
	\subfloat{%
		\includegraphics[height=4.5cm,width=.33\textwidth]{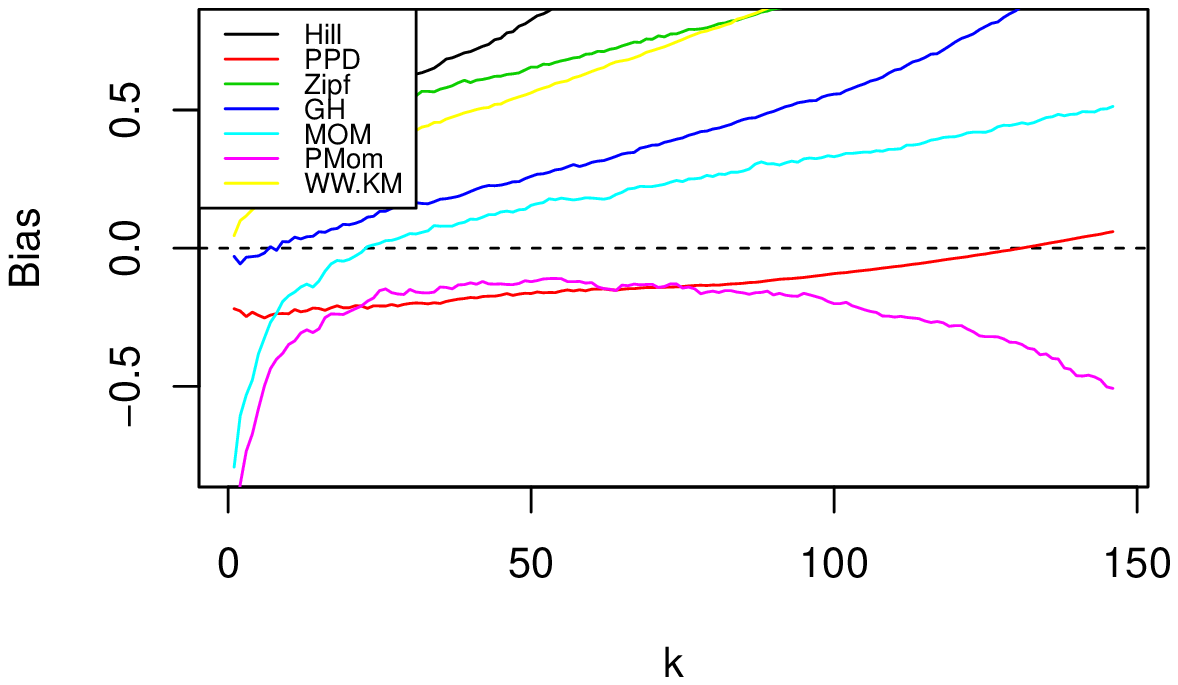}}%
	\subfloat{%
		\includegraphics[height=4.5cm,width=.33\textwidth]{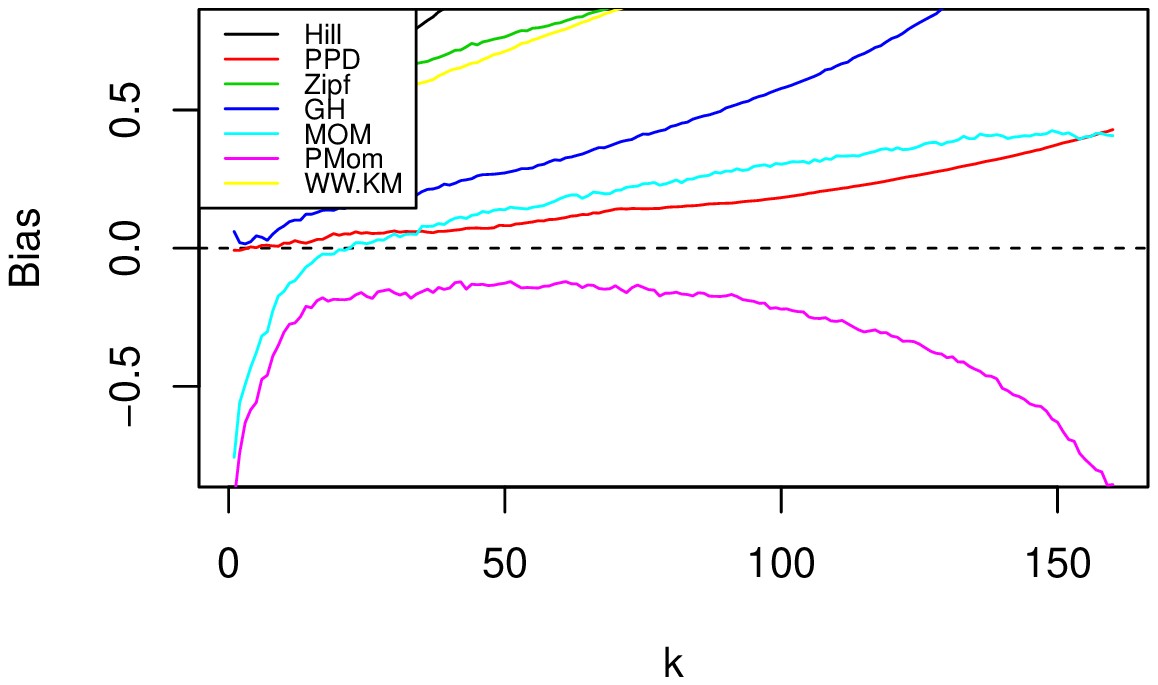}}%
	\subfloat{%
		\includegraphics[height=4.5cm,width=.33\textwidth]{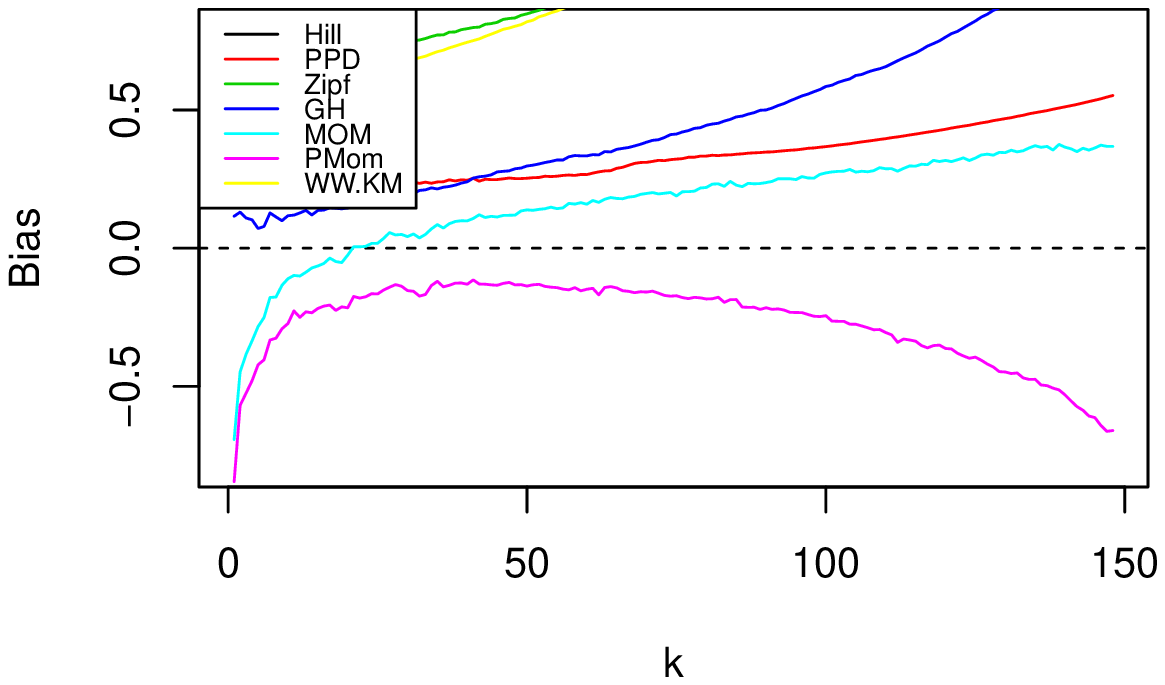}}\\
	
	\caption{ Results for Pareto distribution with $\wp=0.55.$ }
	\label{}
\end{figure}

\section{ Fr\'{e}chet Distribution}
For each figure, the following description of the panels apply.  Leftmost column: $\gamma_1(x)=0.63~ (x=0.12);$ Middlemost column : $\gamma_1(x)=0.31~ (x=0.37);$ Rightmost column: $\gamma_1(x)=0.10~ (x=0.75);$	
	\begin{figure}[H]
		\centering
		\subfloat{%
			\includegraphics[height=4.5cm,width=.33\textwidth]{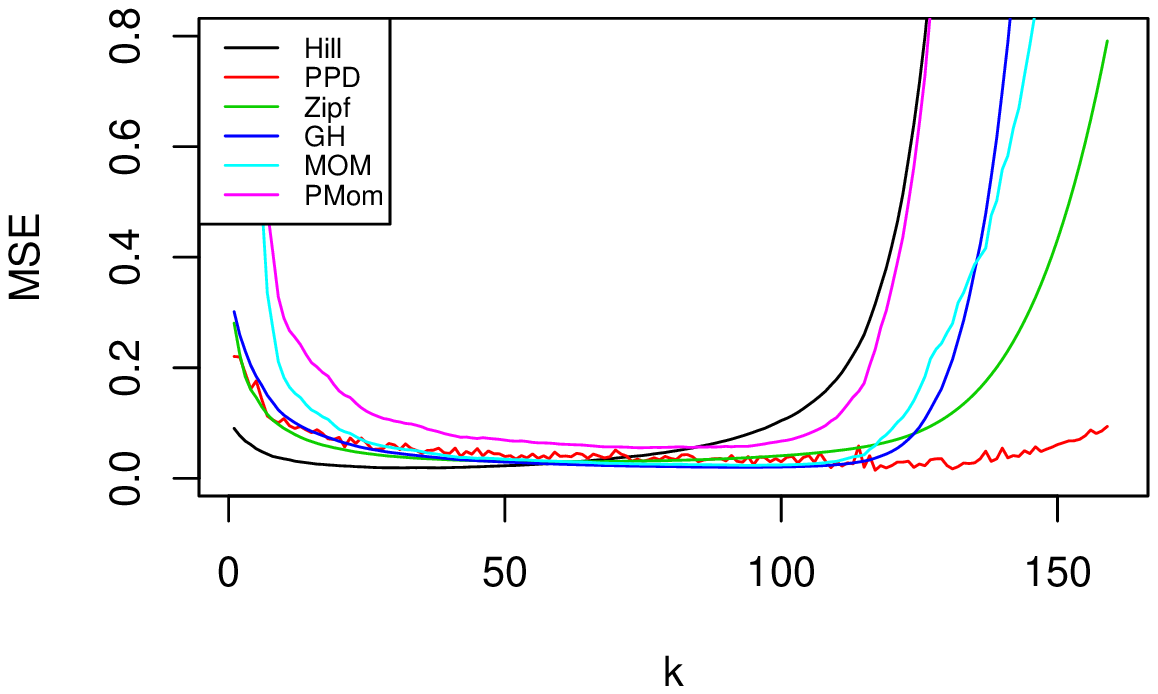}}%
		\subfloat{%
			\includegraphics[height=4.5cm,width=.33\textwidth]{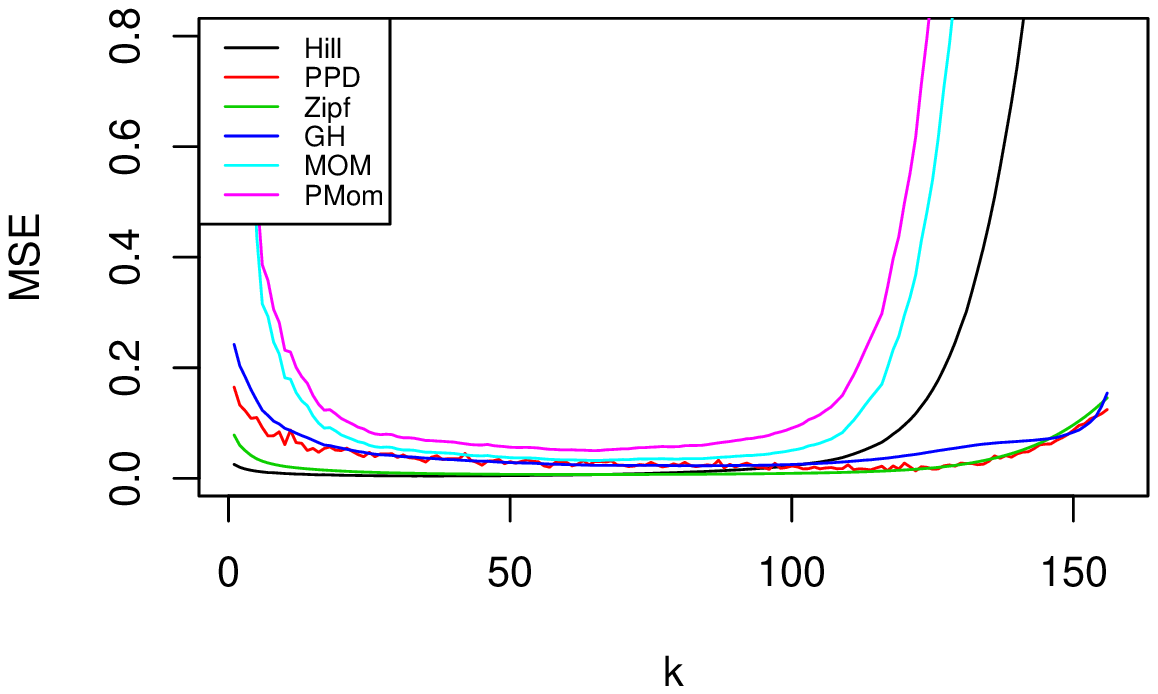}}%
		\subfloat{%
			\includegraphics[height=4.5cm,width=.33\textwidth]{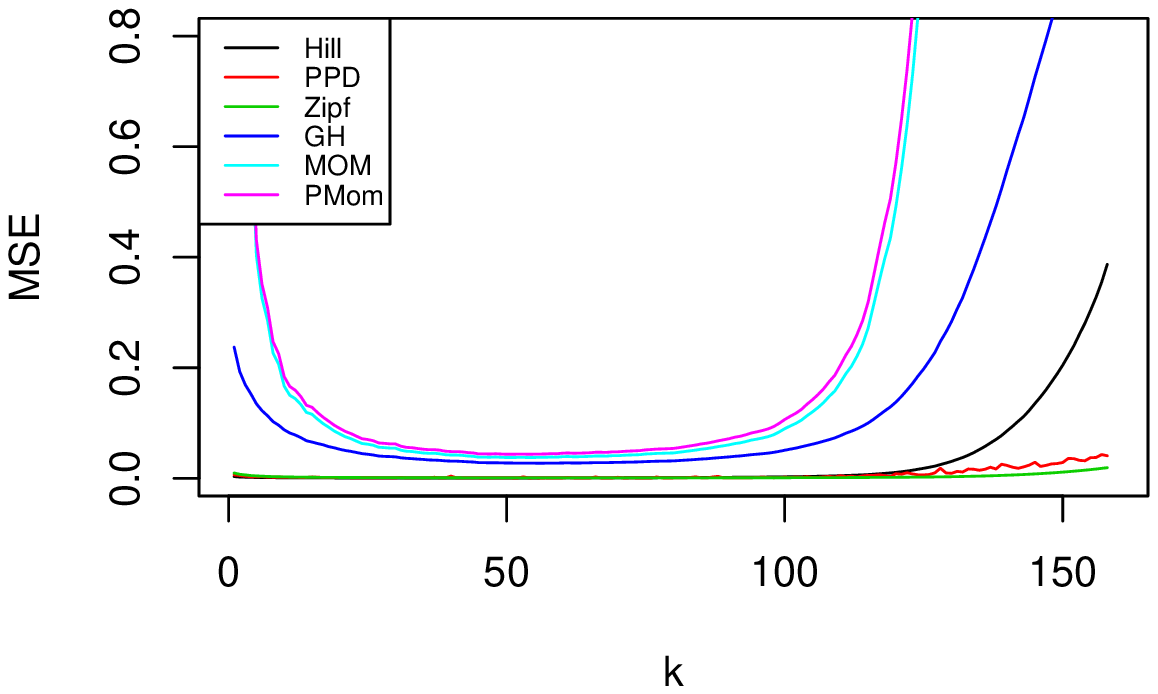}}\\
		\subfloat{%
			\includegraphics[height=4.5cm,width=.33\textwidth]{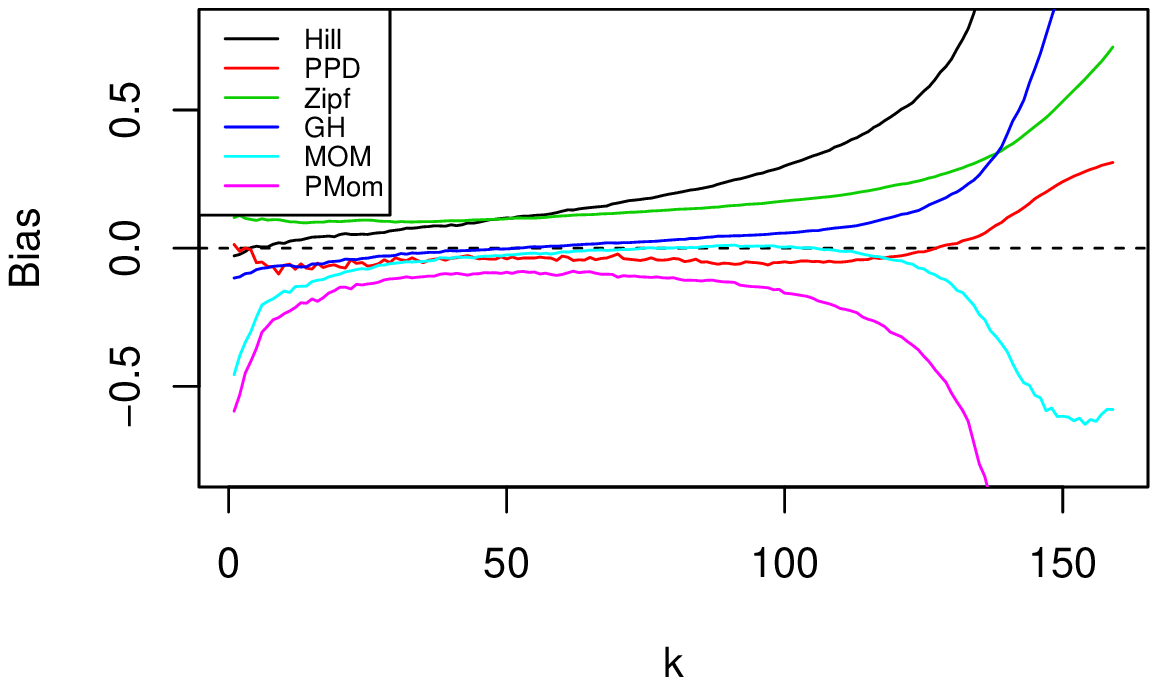}}%
		\subfloat{%
			\includegraphics[height=4.5cm,width=.33\textwidth]{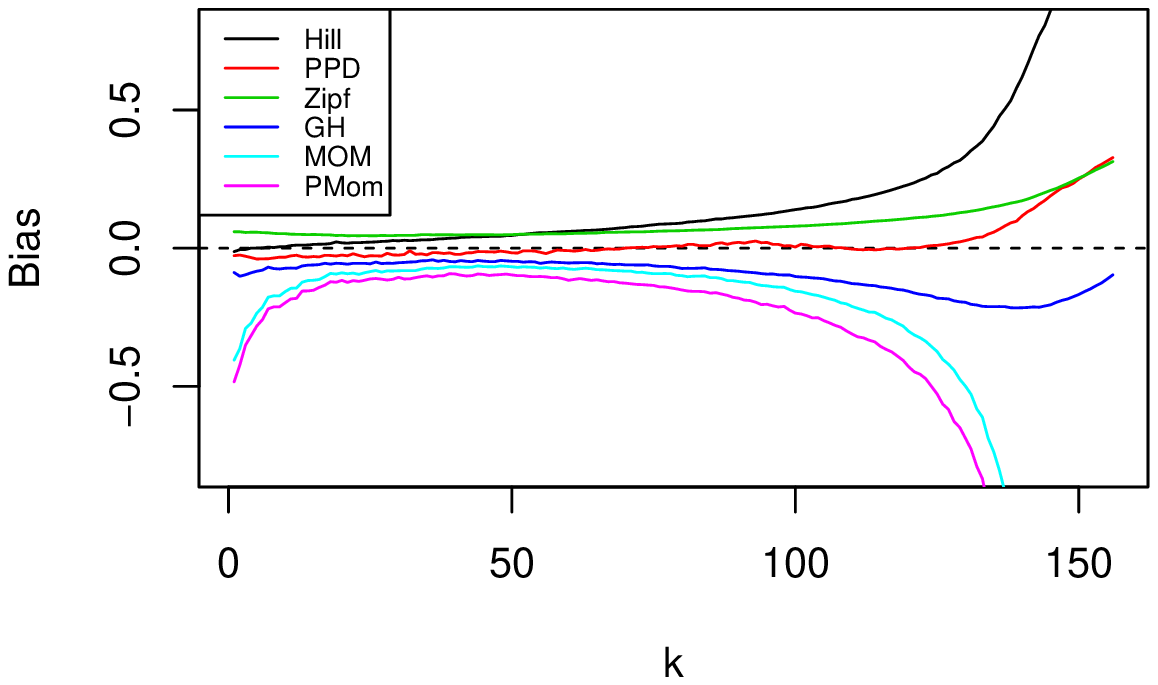}}%
		\subfloat{%
			\includegraphics[height=4.5cm,width=.33\textwidth]{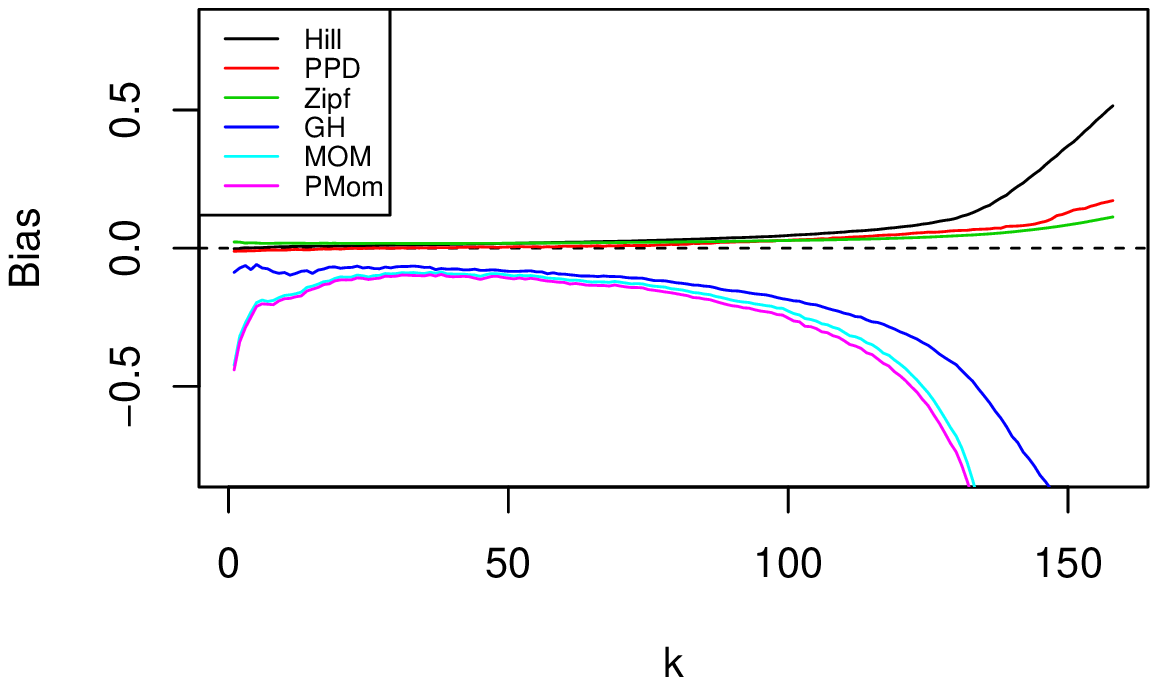}}\\
		
		\caption{ Results for Fr\'{e}chet distribution with $\wp=0.1.$ }
		\label{}
	\end{figure}

	
	\begin{figure}[H]
		\centering
		\subfloat{%
			\includegraphics[height=4.5cm,width=.33\textwidth]{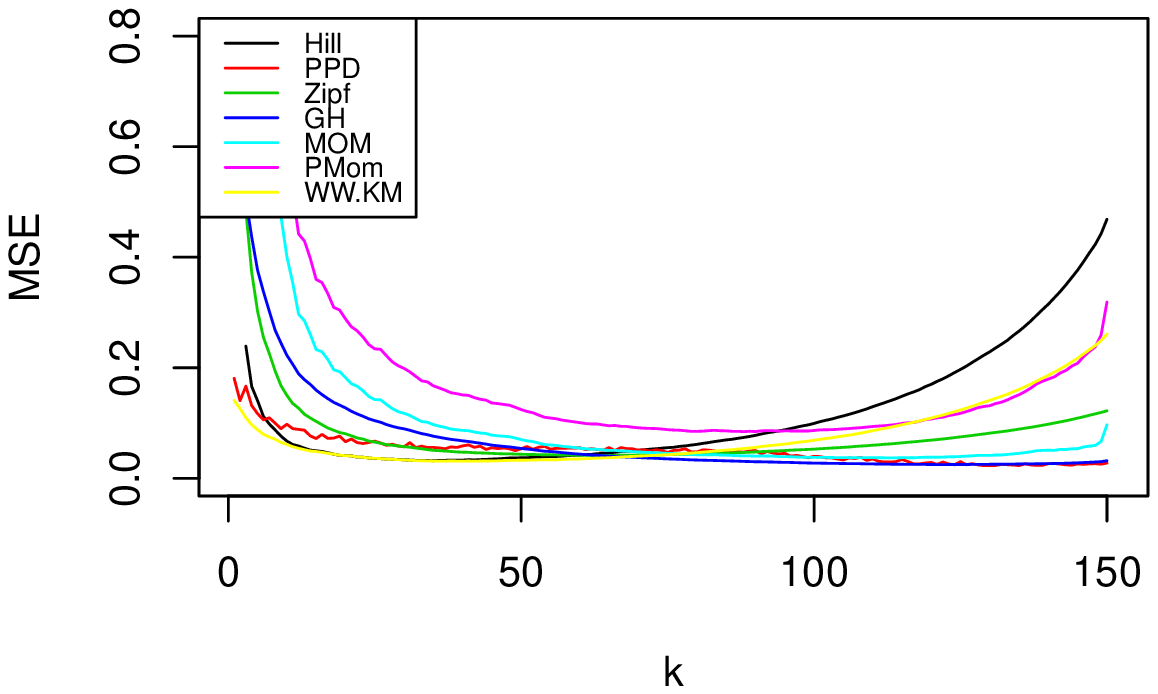}}%
		\subfloat{%
			\includegraphics[height=4.5cm,width=.33\textwidth]{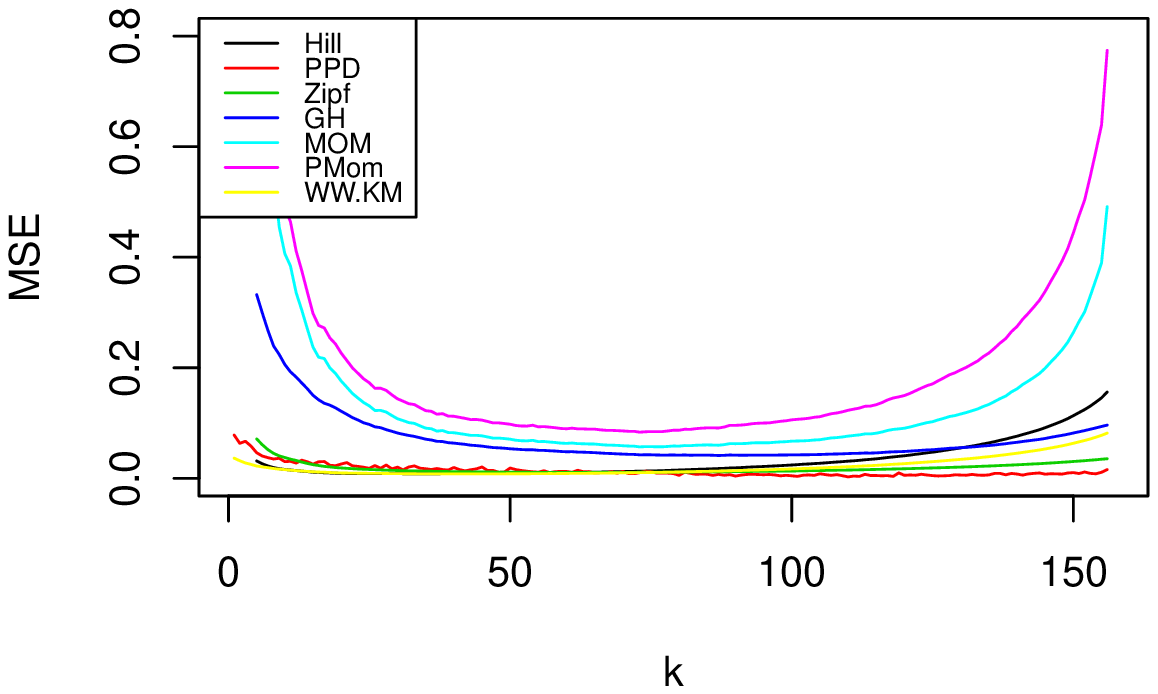}}%
		\subfloat{%
			\includegraphics[height=4.5cm,width=.33\textwidth]{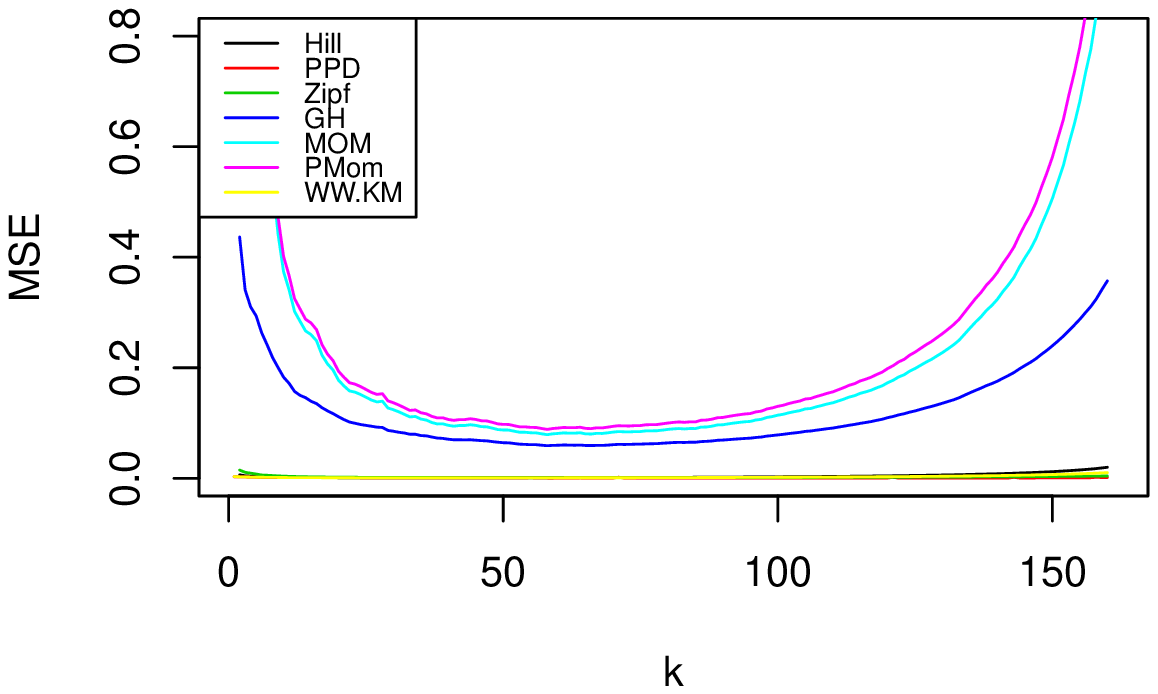}}\\
		\subfloat{%
			\includegraphics[height=4.5cm,width=.33\textwidth]{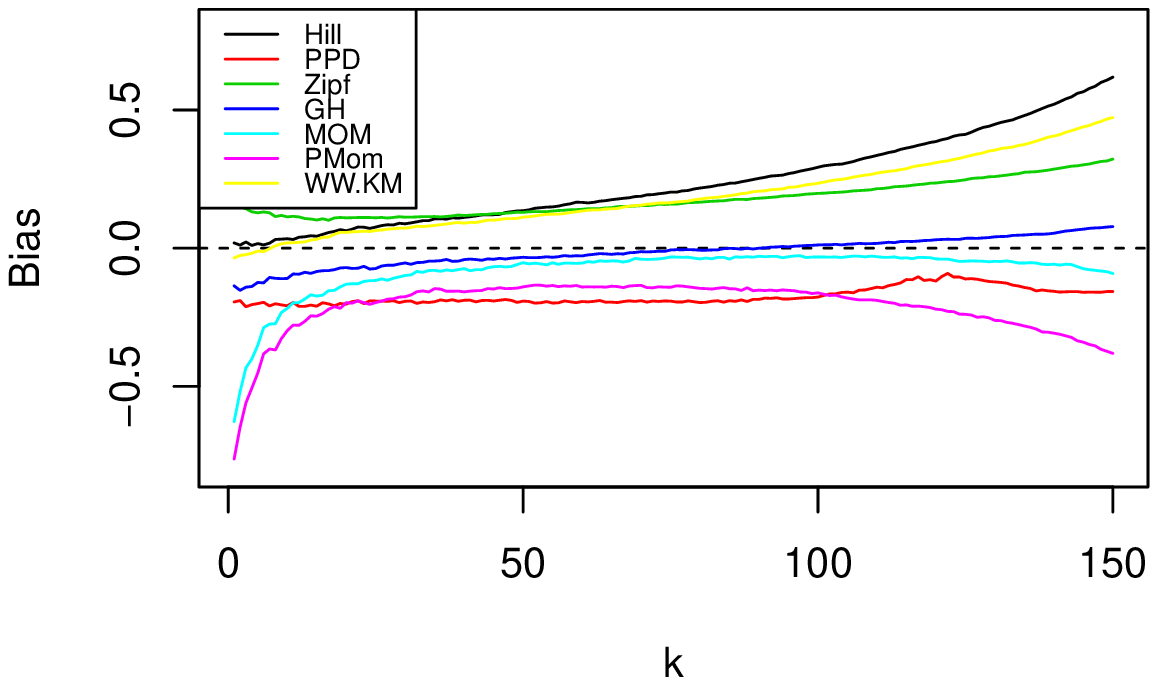}}%
		\subfloat{%
			\includegraphics[height=4.5cm,width=.33\textwidth]{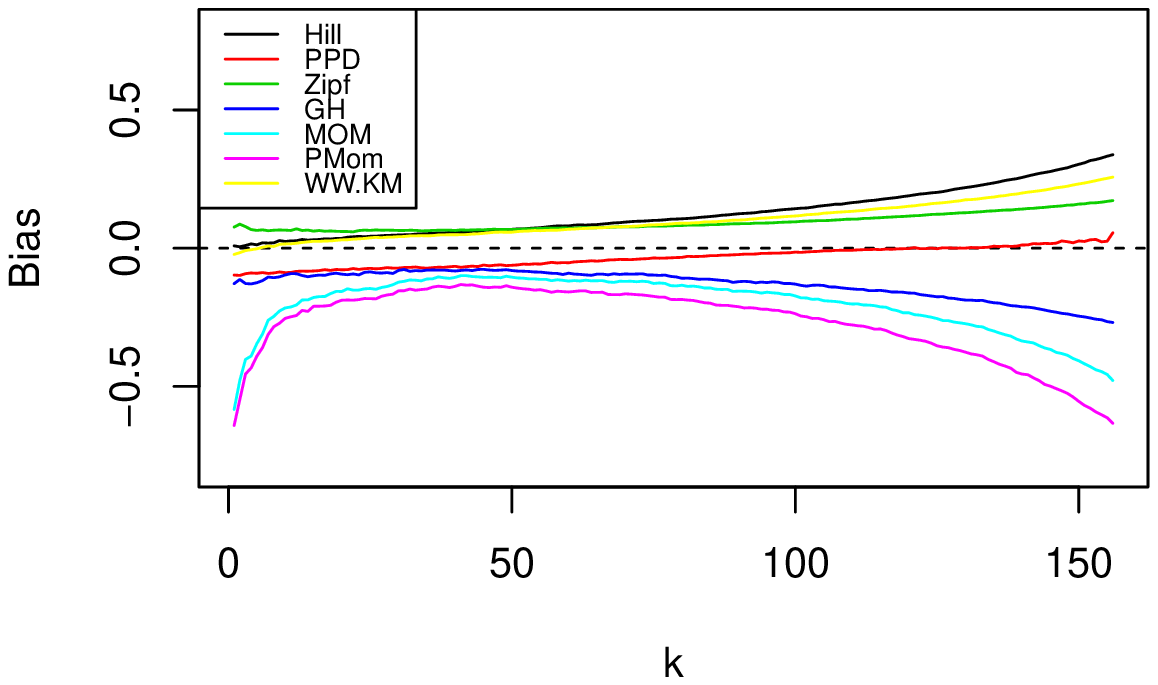}}%
		\subfloat{%
			\includegraphics[height=4.5cm,width=.33\textwidth]{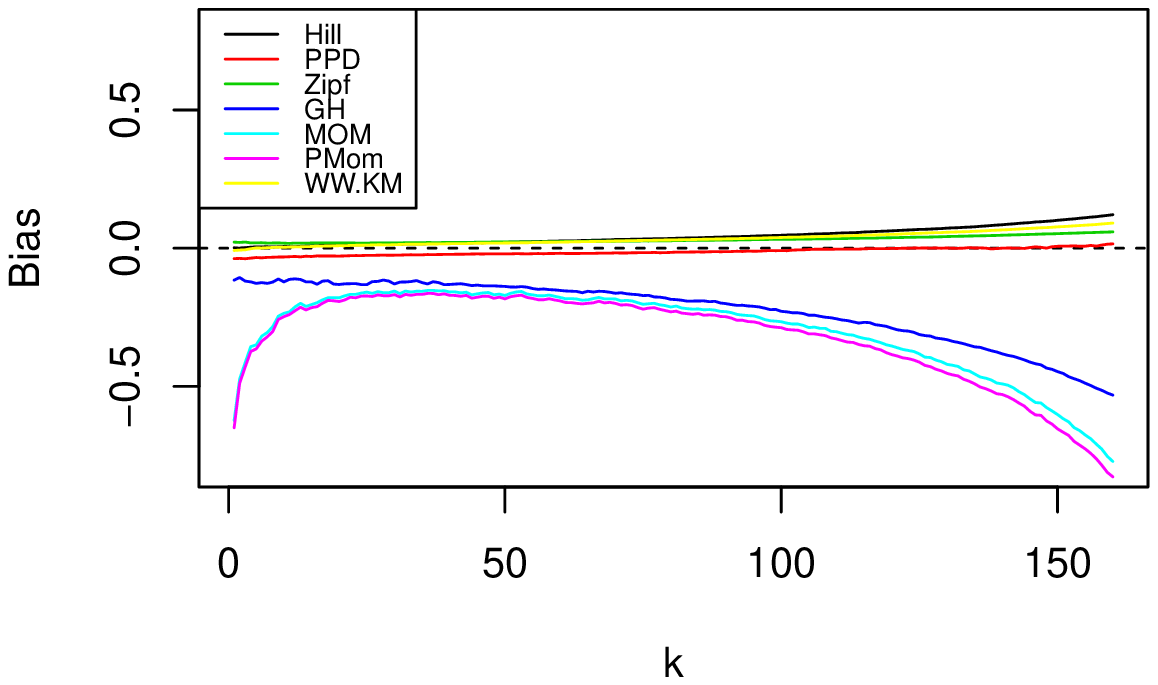}}\\
		
		\caption{ Results for Fr\'{e}chet distribution with $\wp=0.35.$ }
		\label{}
	\end{figure}

	\begin{figure}[H]
		\centering
		\subfloat{%
			\includegraphics[height=4.5cm,width=.33\textwidth]{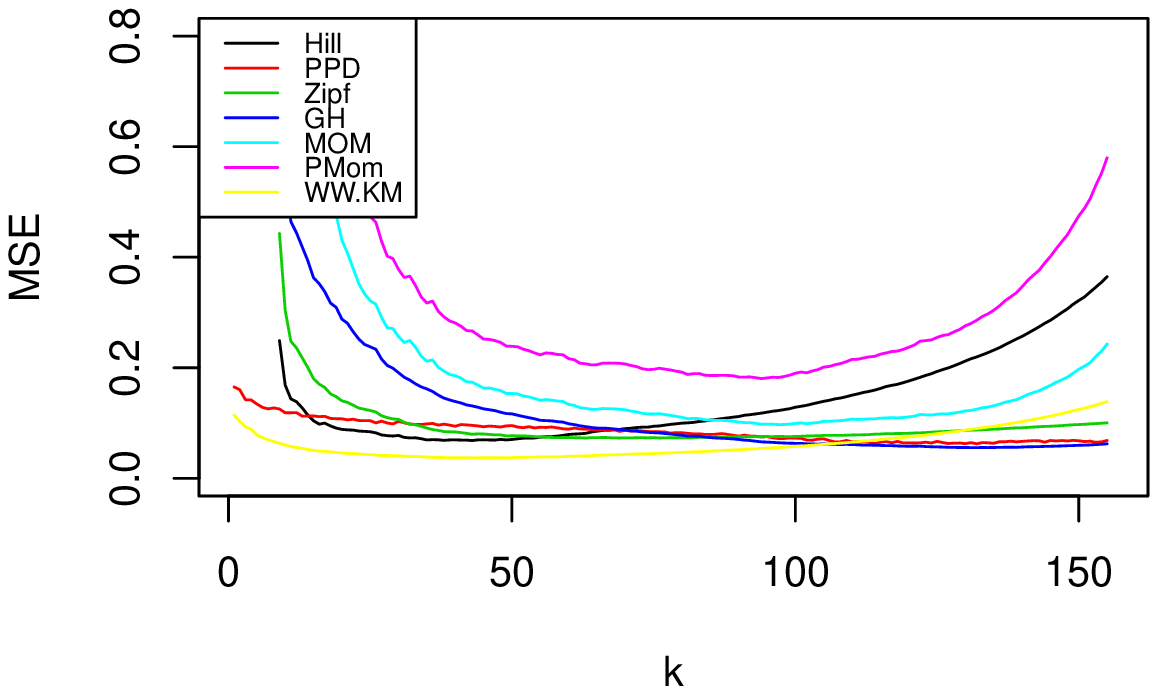}}%
		\subfloat{%
			\includegraphics[height=4.5cm,width=.33\textwidth]{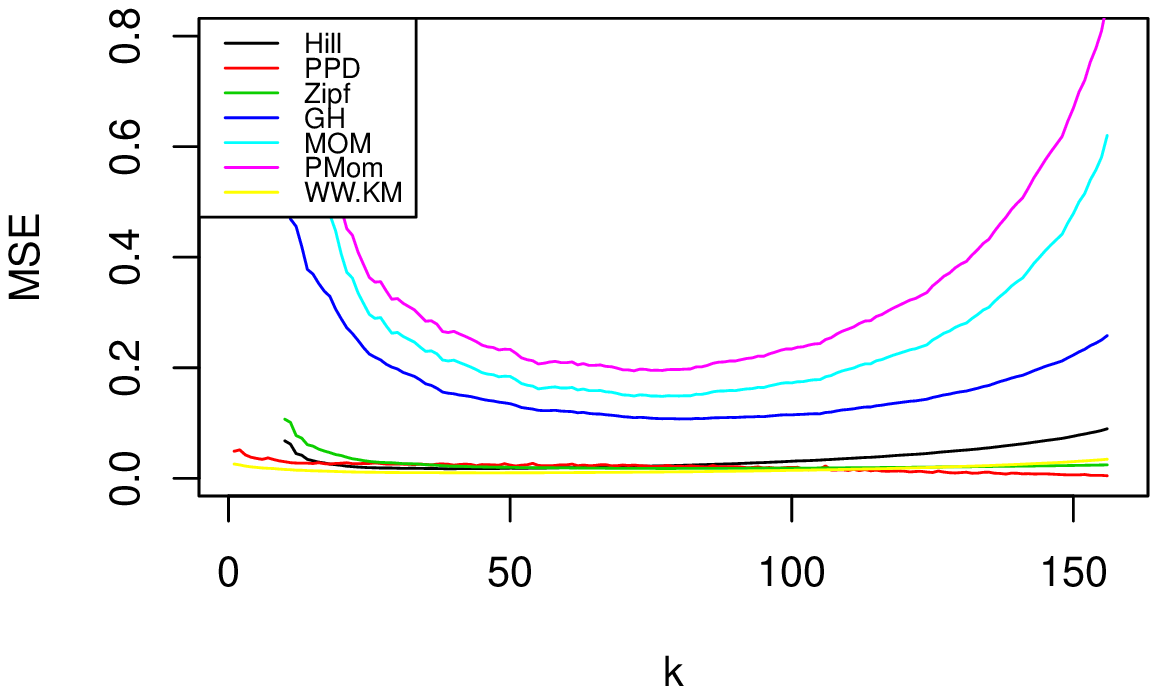}}%
		\subfloat{%
			\includegraphics[height=4.5cm,width=.33\textwidth]{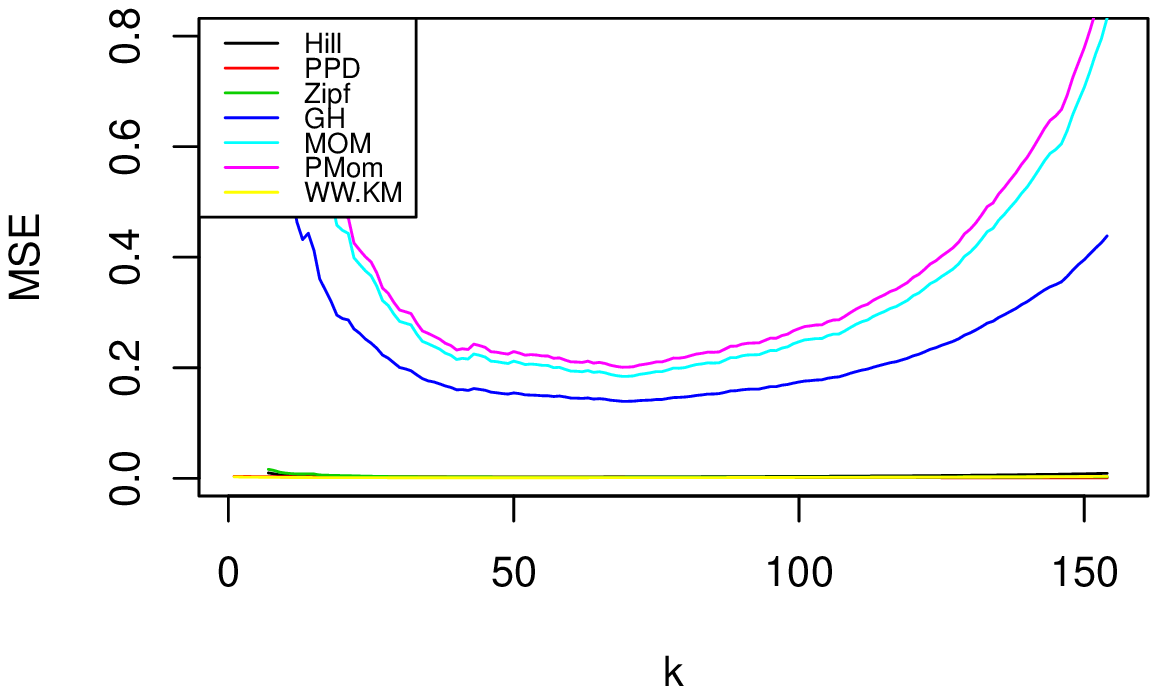}}\\
		\subfloat{%
			\includegraphics[height=4.5cm,width=.33\textwidth]{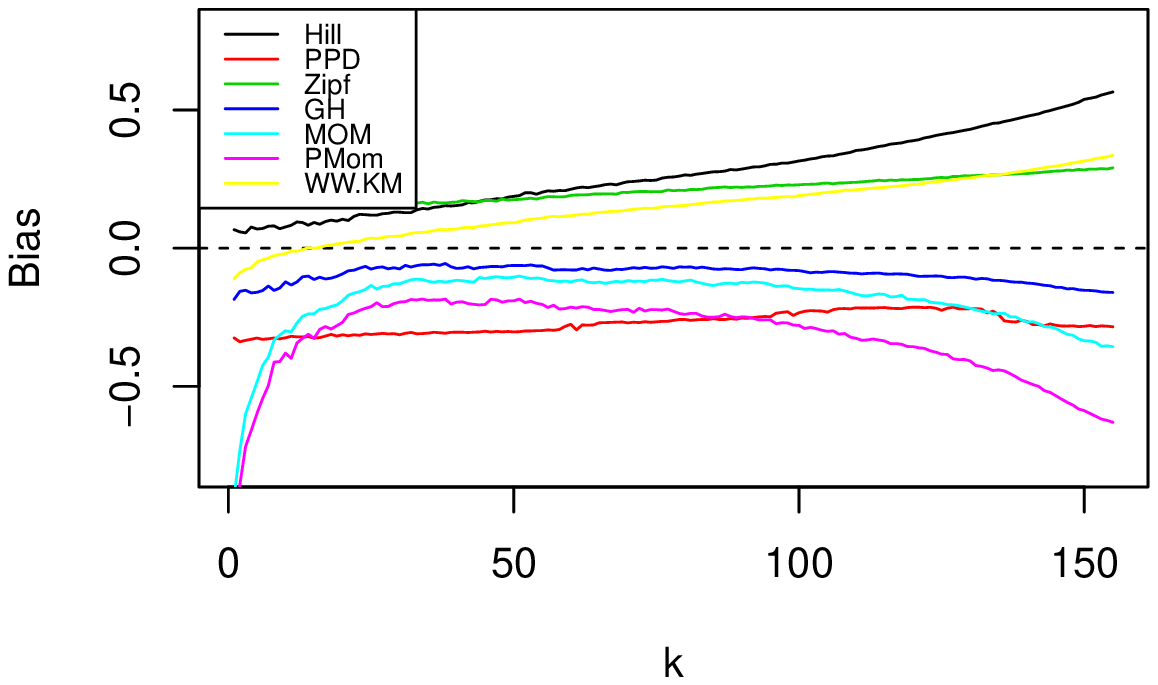}}%
		\subfloat{%
			\includegraphics[height=4.5cm,width=.33\textwidth]{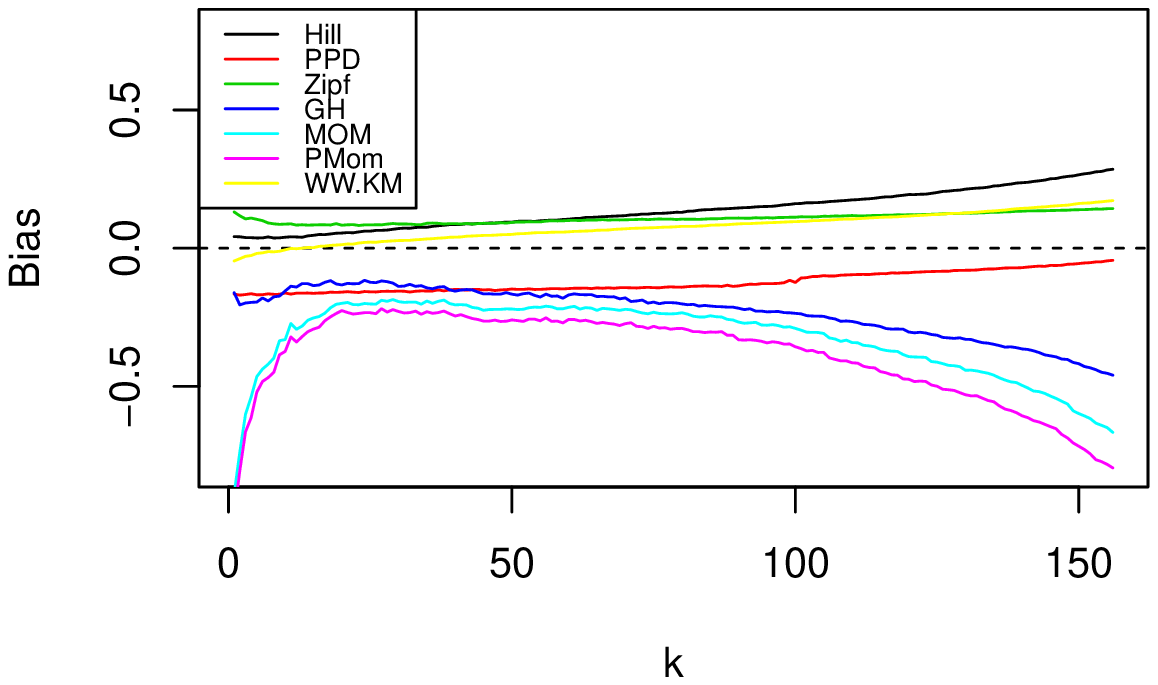}}%
		\subfloat{%
			\includegraphics[height=4.5cm,width=.33\textwidth]{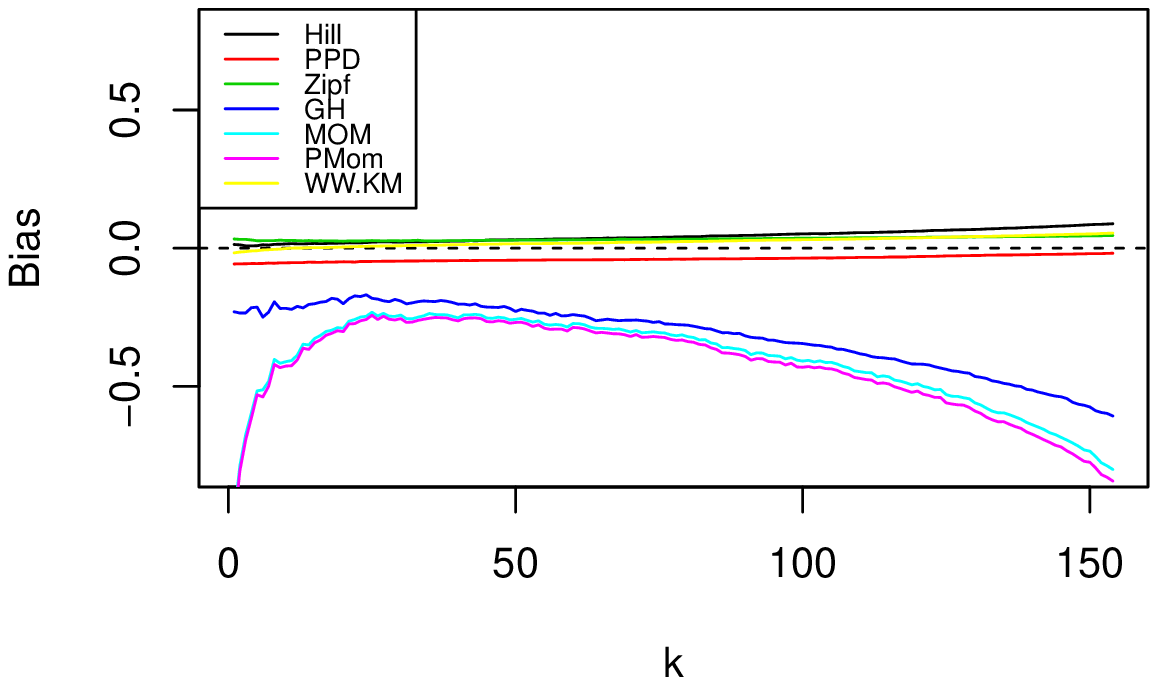}}\\
		
		\caption{ Results for Fr\'{e}chet distribution with $\wp=0.55.$}
		\label{}
	\end{figure}

\end{appendices}
\end{document}